\documentclass[aps]{revtex4}
\usepackage{graphicx}
\usepackage{bm}
\newtheorem{theorem}{Theorem}
\newtheorem{definition}{Definition}
\newtheorem{remark}{Remark}

\newcommand{\middlefig}{.5\textwidth}

\DeclareSymbolFont{AMSb}{U}{msb}{m}{n}
\DeclareSymbolFontAlphabet{\mathbb}{AMSb}
\newcommand{\R}{{\mathbb{R}}}
\newcommand{\C}{{\mathbb{C}}}

\newcommand{\Z}{{\mathbb{Z}}}
\newcommand{\abs}[1]{\vert#1\vert}
\newcommand{\Abs}[1]{\left\vert#1\right\vert}
\newcommand{\norm}[1]{\Vert#1\Vert}
\newcommand{\Norm}[1]{\left\Vert#1\right\Vert}
\newcommand{\At}[1]{\biggr\vert\sb{\sb{#1}}}

\newcommand{\p}{\partial}
\renewcommand{\wp}{\omega_{\mathrm{p}}}

\newcommand\sothat{{\rm;}\ }

\begin{document}

\title{Discrete peakons}

\author{
A. Comech$^1$,
J. Cuevas$^2$,
and P.G. Kevrekidis$^3$
}
\affiliation{
$^3$ Mathematics Department, Texas A\&M University,
College Station, TX 77843-3368, USA,}
\affiliation{$^2$ Grupo de F\'{\i}sica No Lineal, Departamento de
F\'{\i}sica Aplicada I, ETSI Inform\'{a}tica, Universidad de
Sevilla, Avda. Reina Mercedes, s/n. 41012-Sevilla, Spain,}
\affiliation{$^1$ Department of Mathematics and Statistics, University
of Massachusetts, Amherst, MA 01003-4515, USA}

\begin{abstract}
We demonstrate the possibility for
explicit  construction in a discrete Hamiltonian model of an
exact solution
of the form $\exp(-|n|)$, i.e., a discrete peakon. These discrete analogs
of the well-known, continuum peakons of the Camassa-Holm equation
[Phys. Rev. Lett. {\bf 71}, 1661 (1993)] are found in a model
different from their continuum siblings and from earlier studies
in the discrete setting [Phys. Rev. Lett. {\bf 83}, 248 (1999)].
Namely, we observe discrete peakons
in Klein-Gordon-type
and nonlinear Schr\"odinger-type chains
with long-range interactions.
The interesting
linear stability differences between these two chains
are examined numerically and illustrated analytically.
Additionally, inter-site centered peakons are also obtained in explicit
form and their stability is studied.
We also prove the global well-posedness for the discrete
Klein-Gordon equation,
show the instability of the peakon solution,
and the possibility of a formation of a {\it breathing} peakon.
\end{abstract}

\maketitle

\section{Introduction}
\label{sec:introduction}

In the last two decades, intrinsic localized modes (ILMs), also termed
discrete breathers
(DBs), have become a topic of intense theoretical
and experimental investigation; see, e.g., \cite{review} for a number of
recent reviews on the topic. Per their inherent ability to bottleneck
and potentially transport the energy in a coherent fashion,
such exponentially localized in space and periodic in time entities
have come to be of interest in a variety of contexts. These range
from nonlinear optics and arrays of waveguides \cite{mora} to Bose-Einstein
condensates (BECs) inside optical lattice potentials \cite{tromb} and from
prototypical models of nonlinear springs \cite{pesa} to Josephson
junctions \cite{alex} and dynamical models of the
DNA double strand \cite{Peybi}.

The ubiquitous nature of nonlinear lattice systems (i.e., arrays
of coupled nonlinear oscillators) has prompted the examination of
the behavior of nonlinear waves in these systems, particularly
for the waves that are well-known and understood in the continuum
analogs of these equations i.e., in nonlinear partial differential
equations. In this direction of activity, many of the coherent,
nonlinear wave structures that have previously been discovered
in continuum settings such as regular solitons \cite{malomed}, compactons
\cite{rosenau}, shock waves \cite{shocks} or
gap solitons \cite{chen} have been recently examined
in the discrete setting; see, e.g., the works of \cite{review} for
regular solitary waves, \cite{johan} for discrete compactons,
\cite{konotop} for discrete shock waves or
\cite{andrey} for lattice gap solitons.

On the other hand, a continuum nonlinear wave that was recently
obtained in the context of shallow water wave equations, namely
the peakon (a peaked soliton solution with a discontinuity in
the first derivative at its peak) only very recently started to
be considered in the discrete context\cite{oster}.
Peakons were given their name (in the context of the so-called
Camassa-Holm equation, which is a dispersive, integrable model \cite{ch1,ch2})
due to their discontinuous derivative at peak amplitude. It is
worth noting, however, that such ``sharply pointed'' structures were
also obtained much earlier in the context of nonlocal models for
plasmas (see, e.g., \cite{litvak}).
While, at first glance, it may not appear particularly natural to have
discrete analogs of these solutions, as  derivatives are
not, strictly speaking, defined in the context of the spatial
lattice, our aim in the present work is to examine the ``discrete peakon''.
This will be a lattice profile in the form (in fact, exactly)
$\sim \exp(-a |n|)$ that in the continuum limit will asymptotically approach
the continuum peakon solution.

A number of twists accompany this discrete peakon solution
introduced here. In the discussion, we will illustrate that
glimpses of solutions that can be categorized under this new
``species'' may have already appeared in a somewhat generalized
form in earlier works.
We hope this may initiate
a more general examination of the presently suggested lattice peakons.
We
believe that we have encapsulated the key mechanisms whose
interplay can give rise to the existence of peakons (in both the
discrete and the continuum setting) in a model of the Klein-Gordon
(KG) variety and its nonlinear-Schr\"odinger (NLS) type analog.
It is interesting to note that both the dispersive and the
nonlinear terms that we have combined in the present setting have
appeared previously in diverse contexts (that will be mentioned
below); however, they were never combined to allow for the
existence of peakons. Another intriguing bit concerns the
stability properties of the peakons which are also examined in
what follows.
We find that the peakons, while unstable in the KG variant of our
model, become stable in its NLS version.
This is because the negative energy direction present the former model
is prohibited by the charge conservation in the latter model.
In the KG case, discrete {\it breathing} peakons are possible instead.
Interestingly, such solutions were also identified earlier in closed
form in a class of models very different than the ones examined herein
[more specifically in a class of homogeneous, nearest neighbor, Klein-Gordon
Hamiltonians] in \cite{flach}.
Finally, it is
remarkable that the presently examined
class of models and its stability features
can be treated by methods
available for the continuous nonlinear field equations as illustrated
below.

Our presentation will be structured as follows. In Section~\ref{sec:model},
we will discuss the model and its motivation. In Section~\ref{sec:numerical},
the numerical results will be presented for site-centered, as well as
for inter-site centered peakons. In Section~\ref{sec:analytical}, we will
examine the stability of such structures
through analytical considerations
based on constrained energy minimization, as well as on functional
analytic arguments. In Section~\ref{sec:breathers}, the existence of discrete
breathing peakons is discussed.
Finally, in Section~\ref{sec:discussion},
we will summarize our findings and present our conclusions, as
well as some open questions for future studies.

\section{Model and Motivation}
\label{sec:model}

Examining the peakon from the ``reverse engineering'' (or the inverse
problem) point of view, we would like to use the properties of
a solution of the form $u(x) \sim \exp(-|x|)$, to construct
a (continuum as well as a) discrete lattice with dynamics that
supports such peaked solutions. Adopting this viewpoint, some
of the key properties of $u(x)=\exp(-|x|)$ are that
\begin{eqnarray}
u'(x)=-u(x) {\rm sgn}(x),
\label{jeq1}
\\
(1-\partial\sb{x}^2)u(x)=2 \delta(x) u(x).
\label{jeq2}
\end{eqnarray}
This second property (cf. also \cite{ch1})
justifies why a strongly localized impurity (of the form of a $\delta$-function)
may be a relevant context in which such peaked solutions could
arise.
We will return to this point in Section~\ref{sec:discussion}.

Another, perhaps more interesting for our purposes, property is
the result of convolution of such a peaked function
with a Kac-Baker exponential interaction kernel $J(|x-y|)=\exp(-|x-y|)$ \cite{kac,baker}.
The convolution yields:
\begin{eqnarray}
J \star u \equiv \int\sb{-\infty}^{\infty} \exp \left(-|x-y|\right) \exp(-|y|) dy
= \left(1+|x| \right) \exp\left(-|x|\right). 
\label{jeq3}
\end{eqnarray}

This suggests immediately from a mathematical perspective a Klein-Gordon (KG),
as well as a nonlinear Schr\"odinger (NLS) model with long-range interactions
that would support {\it exact} peakon solutions of the form
$u(x)=A \exp(-a |x|)$.
In particular the KG model would read:
\begin{eqnarray}
u\sb{tt}=a \int\sb{-\infty}^{\infty} \exp(-a |x-y|) u(y) dy -
\left(1 - \frac{1}{2}\ln\left(\frac{u^2}{A^2}\right) \right) u.
\label{jeq4}
\end{eqnarray}
The peakons in this case would represent static solutions of the KG
equation. The corresponding NLS model would be of the form
\begin{eqnarray}
i u\sb t= -a \int\sb{-\infty}^{\infty} \exp(-a |x-y|) u(y) dy -
\frac{1}{2} \ln \left(\frac{|u|^2}{A^2}\right) u,
\label{jeq5}
\end{eqnarray}
wherein the peakons correspond to standing wave solutions of
the form $u(x)=A \exp(i t) \exp(-a |x|)$.

Continuing along this lane of reverse construction (the motivation
for each of the relevant terms will be given below), we now explain
that an interesting feature of the above considerations is that they
can also be carried through in the discrete setting. In particular,
for $u\sb n=\exp(-a |n|)$,
we sum up the geometric series establishing that:
\begin{eqnarray}
\sum\sb{m \in \Z}
\exp(-a |n-m|) u\sb m=
\left(\frac{\exp(2 a)+1}{\exp(2a)-1}  +|n|\right) u\sb n.
\label{jeq6}
\end{eqnarray}
Consequently, for the discrete Kac-Baker interaction kernel
\begin{eqnarray}
J\sb{nm}=
\exp(-|n-m|),
\label{extra}
\end{eqnarray}
 we can devise a discrete KG, as well as a discrete NLS
model that have discrete analogs of peaked solutions.
The discrete KG model is of the form:
\begin{eqnarray}
\ddot{u}\sb n=\sum\sb{m \in \Z} J\sb{nm} u\sb m -
\left[\frac{\exp(2 a)+1}{\exp(2a)-1} -\frac{1}{2 a} \ln\left(\frac{u\sb n^2}{A^2}\right) \right]
u\sb n,
\label{jeq7}
\end{eqnarray}
with the {\it exact} discrete peakon solution
$\pi\sb n=A \exp(-a |n|)$,
while the corresponding discrete NLS chain can be formulated as:
\begin{eqnarray}
i \dot{u}\sb n=-\sum\sb{m \in \Z} J\sb{nm} u\sb m + \left[\frac{2}{\exp(2a)-1}- \frac{1}{2a}
\ln\left(\frac{|u\sb n|^2}{A^2}\right) \right] u\sb n,
\label{jeq8}
\end{eqnarray}
where a discrete peakon given by the standing wave
$\exp(i t)\pi\sb n$ is the exact solution of the model.

It is these latter equations [Eqs.~(\ref{jeq7}) and (\ref{jeq8})]
and their dependence on the parameter $a$
that determines the interaction ``range'' that we plan on investigating
in what follows. Notice that $a$ can be considered
as a natural spacing parameter.

It is interesting to note as a side remark (to which we will return
in later sections) that this model not only supports an exact
``on-site'' discrete peakon solution such as the one given above,
but additionally supports {\it exact} ``inter-site'' peakon solutions
of the form:
$u_n=B \exp(-|n-1/2|)$ for the Klein Gordon (in the DNLS case it
is $u_n=B \exp(i t) \exp(-|n-1/2|)$).
The value of the prefactor $B$ is given by the relation
\begin{equation}
\ln(B/A)=a\left[\frac{(\exp(a)-1)}{2(\exp(a)+1)}\right].
\label{twosite}
\end{equation}
Such explicit solutions (especially inter-site ones) are rarely available
in non-integrable discrete models. Inter-site solutions are typically
less stable than their on-site siblings \cite{ijmpb}.
In the present setting, we will
study in detail the behavior of such two-site peakons numerically
as well as analytically.

In motivating the model, aside from its intrinsic mathematical interest
due to the existence of the peaked solutions (both in the discrete case
and in the continuum limit), we should remark that both the dispersive
and the nonlinear terms included here have appeared in a variety of
settings before. The Kac-Baker type kernel \cite{kac,baker}, aside
from its relevance in models of statistical physics, has been used
quite extensively in recent nonlinear studies of lattice models
emulating biopolymer dynamics including DNA; see, e.g.,
\cite{gaid1,gaid2,gaid3,gaid4,gaid5,gaid6}. Hence, this type of
interaction is rather ubiquitous and can be controllably tuned
(depending on the value of $a$), to be practically nearest neighbor
(for large $a$) or much longer range (for $a \rightarrow 0$).
Let us note also that in the past and in the framework of continuum
equations similar forms of this kernel had been examined in the
work of \cite{witham} (but in a rather different dynamical model,
namely one of the KdV type) wherein it was found that traveling
waves acquired a peaked waveform. In contexts more closely related
to the ones of the present work, let us also mention that
``cusp solitons'' (i.e., peakons) were also found in continuum
models of the NLS type with Kac-Baker interactions in \cite{ybg}
(however, they were unstable)
and in the case of a nonlocal Klein-Gordon field theory in
\cite{egor}.

The logarithmic nonlinearity in nonlinear model
equations was originally introduced in the context of quantum field
theory \cite{rosen}. It reappeared in \cite{bb}, where
it was proposed as an equation for generalized quantum mechanics.
Later in \cite{ms6}, it was suggested as a description for
extended objects in nuclear matter, while more recently it was examined
in the context of a scalar field model in inflationary cosmology
\cite{cosmol}.
These studies have also triggered a more mathematically oriented
interest in this nonlinearity and the properties of the solutions
of the corresponding nonlinear wave equations \cite{maslov}. Perhaps,
most closely to the purposes of interest to this study, this
type of logarithmic field theory has appeared in saturable nonlinear
optical media. The initial investigations of \cite{sny}
in the latter context, in the framework of the ``mighty morphing'' spatial
solitons, were later placed in a more physically realistic
framework in connection with photorefractive materials in
\cite{christo}. The work of \cite{christo} suggests that
for the  nonlinear waveguide evolution, the logarithmic nonlinearity provides
an accessible model that offers valuable
insight, while maintaining the characteristic features of the
underlying physical process. A note of caution should however be
made in this connection in that the nonlinearity of Eqs.
(\ref{jeq7})-(\ref{jeq8}) should be viewed as a more reasonable
physical model for larger amplitudes (where it can be considered as
an approximation of a more physical nonlinear term such as
${\rm ln}(1+|u_n|^2)$). For amplitudes tending to $0$, the
divergence of ${\rm ln}(|u_n|^2)$ appears to be somewhat unphysical
and leads to the absence of a small amplitude excitation (so-called
``phonon'') spectrum.

The combination of the features of the dispersive interaction
(its controllable range and
wide applicability) and of the logarithmic nonlinearity
(an accessible one representing adequately a number of physical
processes) renders our model a possibly good playground to study,
e.g., an array of coupled saturable nonlinear (logarithmic)
waveguides. Both the potential relevance of our results in this
context, as well as their inherent mathematical interest in
establishing the discrete properties and behavior of the
peaked solutions, lead us to examine Eqs.~(\ref{jeq7}) and
(\ref{jeq8}) in what follows.

\section{Numerical Results}
\label{sec:numerical}

\subsection{General Setup}

The equations that we will examine can be re-written
in a more general form:
\begin{eqnarray}
\ddot{u}\sb n - \sum\sb{m \in \Z}
J\sb{nm}u\sb m + F(u\sb n) = 0
\label{jeq9}
\end{eqnarray}
for the KG lattice and
\begin{eqnarray}
i \dot{u}\sb n=-\sum\sb{m \in \Z}
J\sb{nm}u\sb m  + G(u\sb n,u\sb n^{\star})
\label{jeq10}
\end{eqnarray}
for the DNLS chain; recall that $J \sb{nm}$ is given by Eq.~(\ref{extra}).
For Eqs.~(\ref{jeq7}) and (\ref{jeq8}), the
respective on-site terms are:
\begin{eqnarray}
F(u\sb n)=\left[\frac{\exp(2 a)+1}{\exp(2a)-1} -\frac{1}{2 a}
\ln\left(\frac{u\sb n^2}{A^2}\right) \right] u\sb n,
\label{jeq9a}
\\
G(u\sb n,u\sb n^{\star})=\left[\frac{2}{\exp(2a)-1}- \frac{1}{2a}
\ln\left(\frac{u\sb n u\sb n^{\star}}{A^2}\right) \right] u\sb n.
\label{jeq10a}
\end{eqnarray}
Without loss of generality, we set $A=1$.
The exact solutions of interest
for Eq.~(\ref{jeq9}) are
of the familiar peakon form mentioned previously:
\[
\pi\sb n=\exp(-a|n|).
\]

We examine the linear
stability of these solutions by using
in Eq.~(\ref{jeq9})
\begin{eqnarray}
u\sb n= \pi\sb n + \epsilon \exp(i \omega t) v\sb n,
\label{jeq11}
\end{eqnarray}
where $\pi\sb n$ is the original peakon and $\omega$ are the
eigenfrequencies of linearization around the solution ($v\sb n$ are
the corresponding eigenvectors). The resulting linear stability
equation (obtained by using the ansatz of Eq.~(\ref{jeq11}) to
O$(\epsilon)$ in Eq.~(\ref{jeq9})) reads:
\begin{eqnarray}
-\omega^2 v\sb n = \sum\sb{m \in \Z}
J\sb{nm} v\sb m - F'(\pi\sb n) v\sb n.
\label{jeq12}
\end{eqnarray}
This is an eigenvalue problem
for the matrix
$J\sb{nm}-\delta\sb{n m}F'(\pi\sb n)$.
The discrete peakon is linearly unstable
if there are eigenfrequencies $\omega$
with the negative imaginary part.
Since
the matrix elements $J\sb{n m}$
are bounded and translation-invariant
(only depend on $n-m$)
while
$F'(\pi\sb n)$
exponentially decays as $n\to\infty$,
one can show that
the eigenvalues of the truncated matrix,
with $\abs{m},\,\abs{n}\le N$,
will tend to the eigenvalues of (\ref{jeq12})
as $N\to\infty$.
The eigenvalue for the truncated matrix can easily be solved
using numerical linear algebra packages;
this gives the approximate eigenfrequencies $\omega$
and the corresponding eigenvectors $v\sb n$.

For the DNLS lattice, the stability can be performed in the
``co-rotating'' frame \cite{aubry1}, using the ansatz
\begin{eqnarray}
u\sb n= \exp(i t) \left[\pi\sb n + \epsilon \left( a\sb n \exp(-i \omega t) +
b\sb n^{\star} \exp(i \omega^{\star} t) \right) \right].
\label{jeq13}
\end{eqnarray}
Then, the resulting linear stability equations will be of the
form:
\[
\omega
\left( \begin{array}{c}
a\sb{k} \\
b\sb{k}^{\star} \\
\end{array} \right)
= {\bf J} \cdot
\left( \begin{array}{c}
a\sb{k} \\
b\sb{k}^{\star} \\
\end{array} \right), \\
\]
where ${\bf J}$ is the linear stability (Jacobian) matrix of the
form
\[
{\bf J}=
\left( \begin{array}{cc}
\frac{\partial { {\mathcal F}\sb i}}{\partial u\sb j} & \frac{\partial {{\mathcal F}\sb i}}{\partial u\sb j^{\star}} \\
-\frac{\partial {{\mathcal F}\sb i^{\star}}}{\partial u\sb j} & - \frac{\partial {{\mathcal F}\sb i^{\star}}}{\partial u\sb j^{\star}} \\
\end{array} \right), \\
\]
and ${\mathcal F}\sb n=-\sum\sb{m \in \Z}
J\sb{nm} u\sb m +
G(u\sb n,u\sb n^{\star}) + u\sb n$ (the Jacobian should be evaluated at the peakon profile, $u\sb n=\pi\sb n$).

We now proceed to examine stability
and dynamics properties of peakons in Klein-Gordon and DNLS systems.

\subsection{Klein--Gordon 1-site peakons}

\begin{figure}
\begin{center}
\begin{tabular}{cc}
\hskip -24pt
    \includegraphics[width=\middlefig]{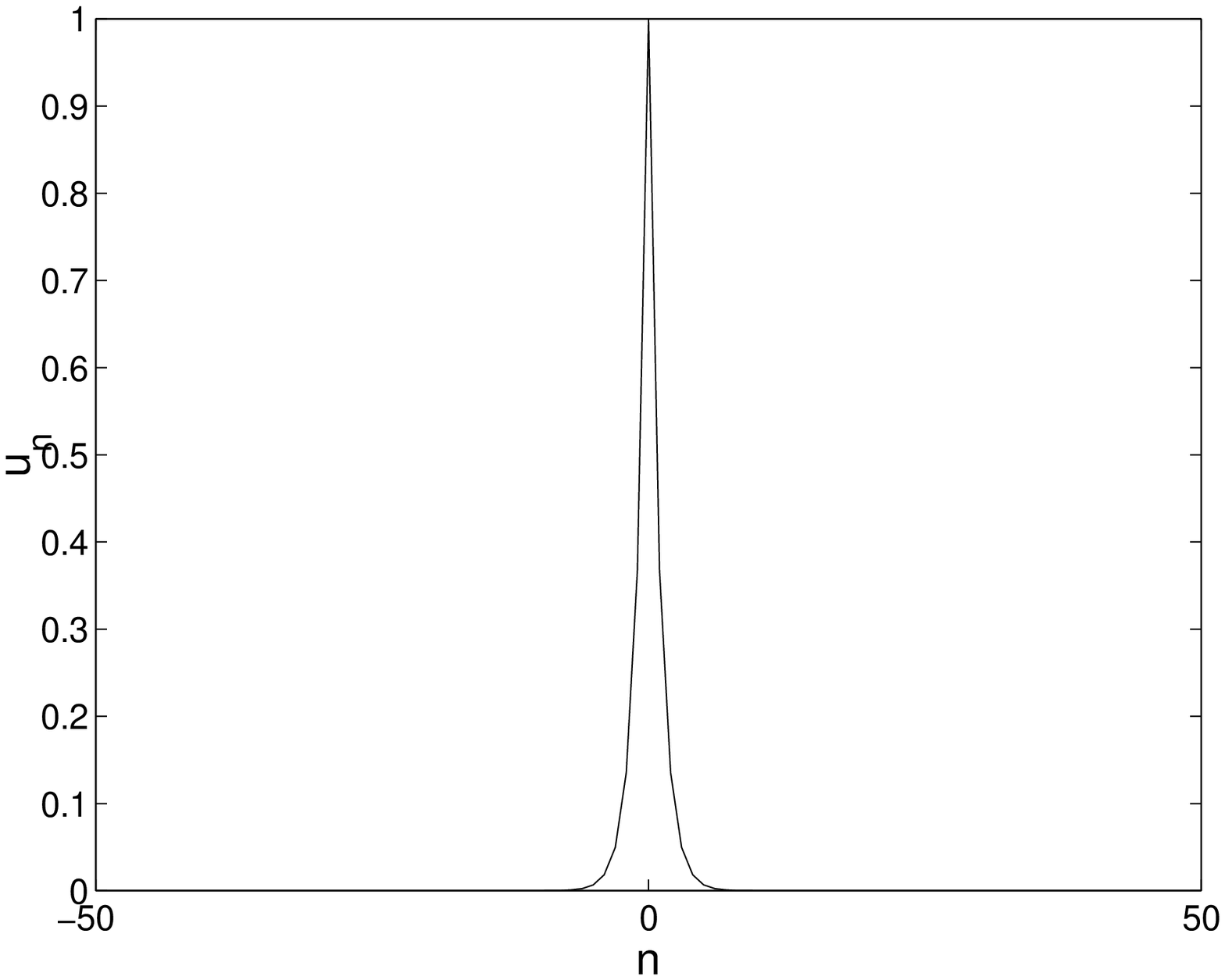} &
    \includegraphics[width=\middlefig]{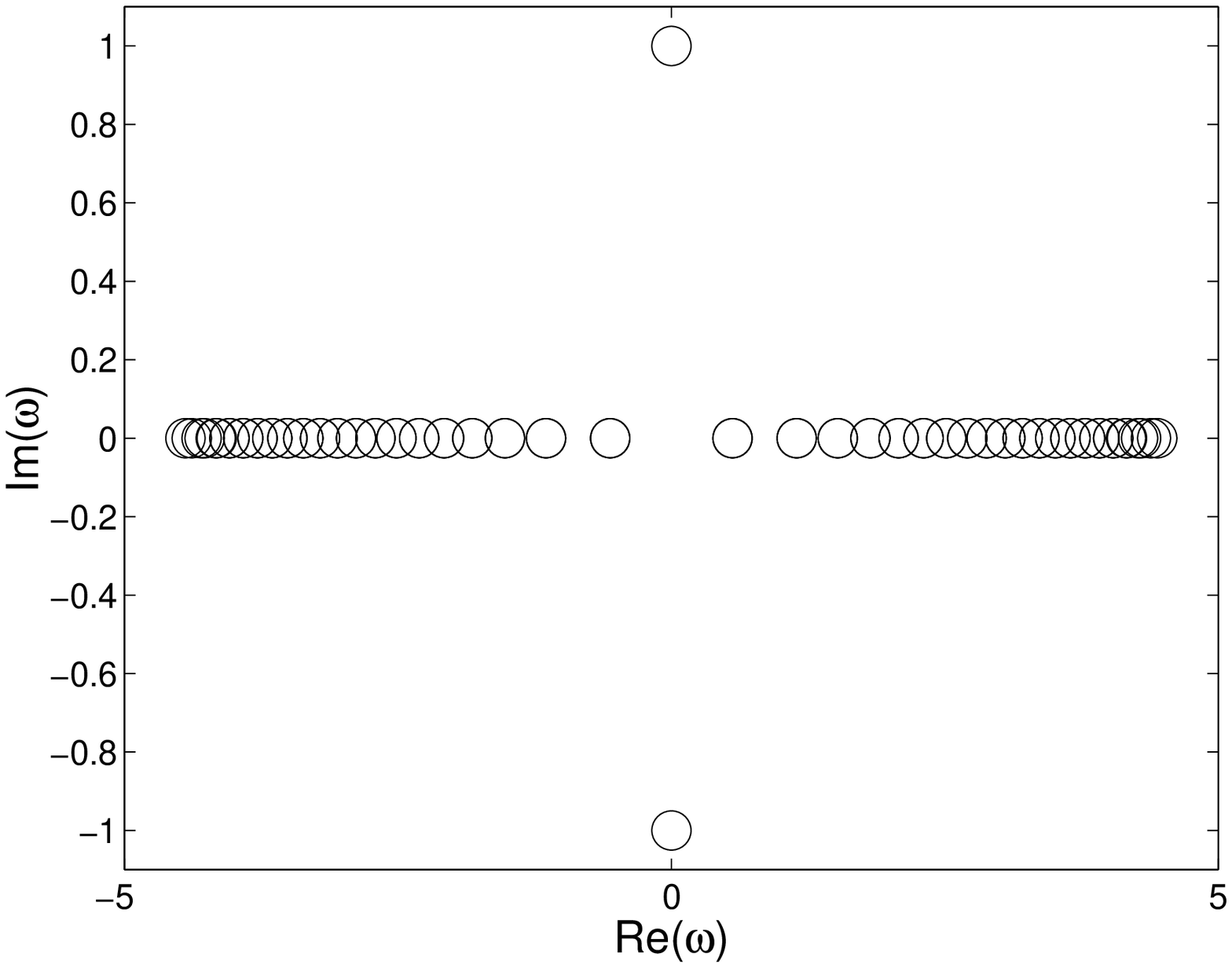}\\
\end{tabular}
\caption{Left panel: Spatial profile of an exact discrete peakon
($a=1$).
\\
Right panel: Spectral plane (Re($\omega$),Im($\omega$))
of the stability matrix for this solution in a Klein--Gordon chain.}
\label{fig:peakonkg}
\end{center}
\end{figure}

The profile of a peakon is shown in Fig.~\ref{fig:peakonkg},
together with the spectrum of eigenfrequencies of the
equation linearized at the peakon.
For the KG chain, we find that the solutions are
unstable (for all values of $a$). This is because of a {\it negative
energy direction} that leads to an imaginary pair of
eigenfrequencies. The
corresponding eigenvector has the same shape as the peakon itself.

\begin{figure}
\begin{center}
\begin{tabular}{cc}
\hskip -24pt
    \includegraphics[width=\middlefig]{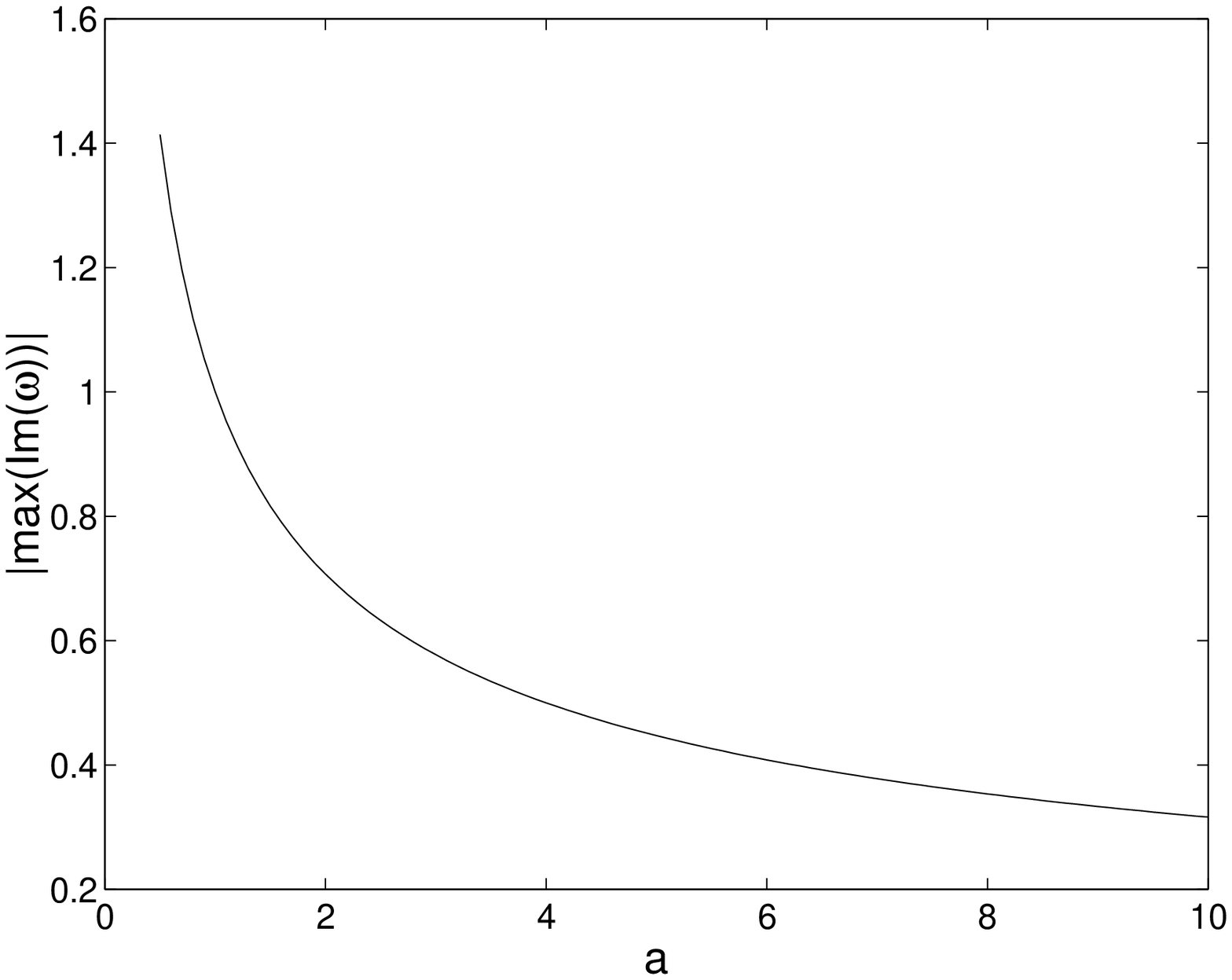} &
    \includegraphics[width=\middlefig]{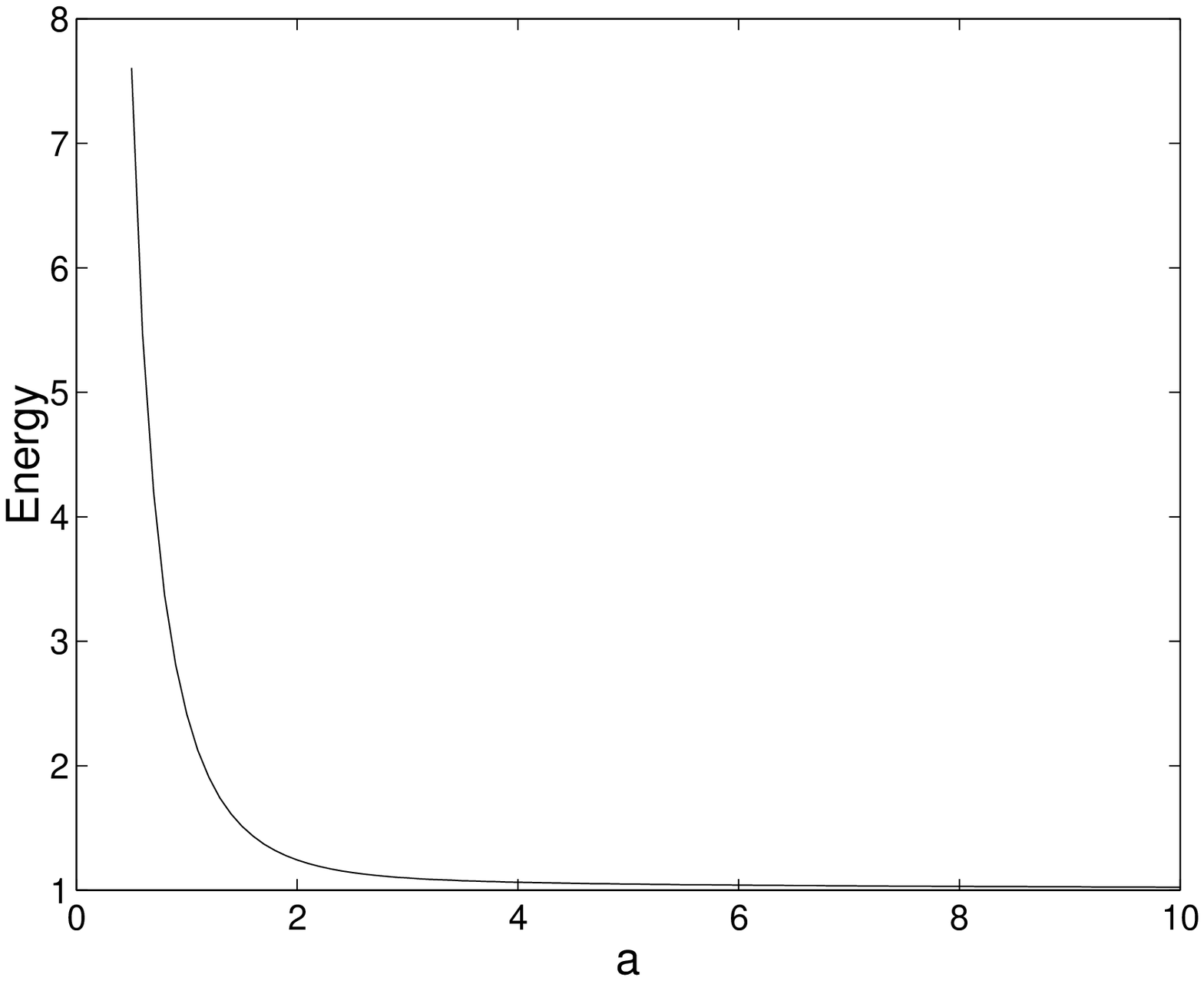}\\
\end{tabular}
\caption{Left panel: Dependence of the maximum imaginary
eigenfrequency (i.e., maximum real eigenvalue)
of the stability matrix on $a$.
\\
Right panel: Dependence
of the energy of the peakon on $a$.}
\label{fig:stabkg}
\end{center}
\end{figure}

This fact can be observed in more detail in Fig.~\ref{fig:stabkg},
where the dependence of the imaginary part of the unstable
eigenvalue of the stability matrix is shown as a function of $a$.
This figure also gives the dependence of the energy of the peakon
on $a$, which can be analytically calculated:
$E(a)=A^2\coth(a)/(2a)$. From these figures, it can be deduced
that the solution becomes less unstable with (increasing) $a$, or,
in other words, when the width of the peakon decreases. This can
be equivalently interpreted as a weaker instability as the
solution approaches its anti-continuum limit, single-site peakon
sibling. This is a rather natural feature of spatial discreteness
which typically serves to stabilize coherent structures that are
unstable in the continuum limit (e.g., due to collapse) or even
ones that do not exist in that limit \cite{review}.

\begin{figure}
\begin{center}
\begin{tabular}{cc}
    \includegraphics[width=\middlefig]{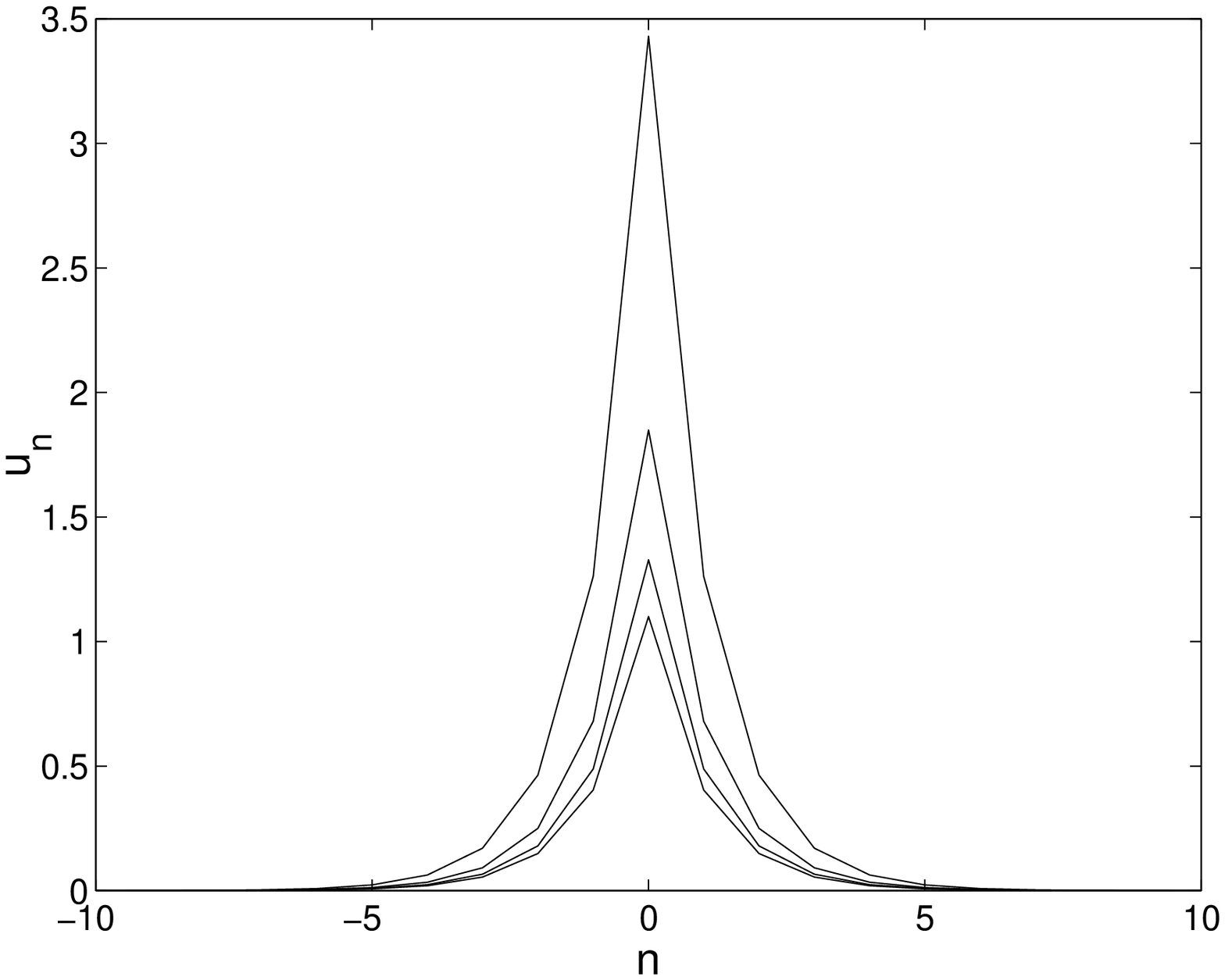} &
    \includegraphics[width=\middlefig]{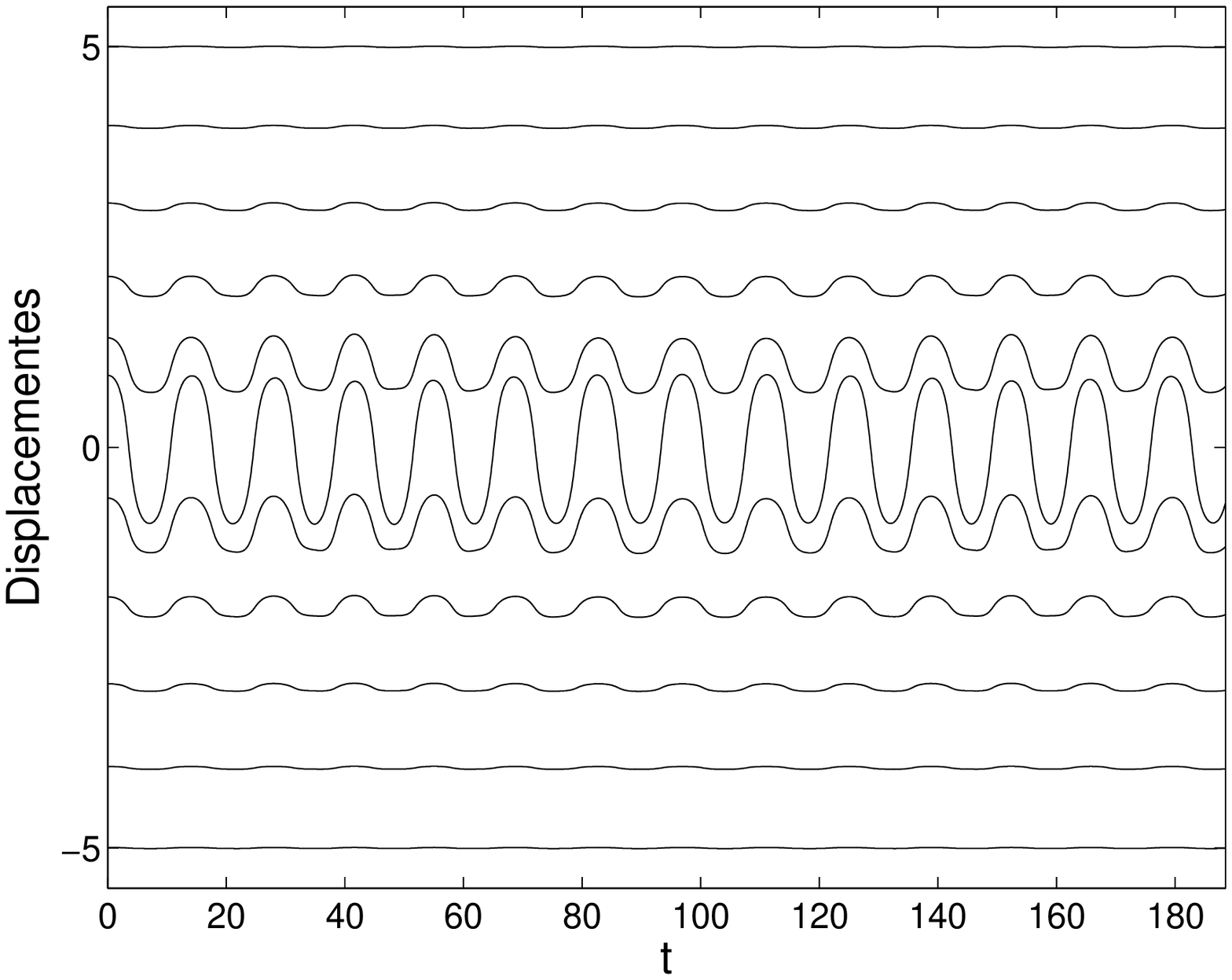} \\
\end{tabular}
\caption{Left panel: Instability evolution of a 
discrete peakon at different times ($a=1$).
Exponential growth can be observed, while the energy of the system
is preserved up to a $10^{-15}$ precision. The different snapshots
of the solution correspond (from inner to outer) to times 0, 1.8,
2.7 and 3.6.
\\
Right panel: Time evolution of a perturbed peakon that develops
into a breathing state. The displacement of the central sites of
the peakon are shown as a function of time.}
\label{fig:simulkg}
\end{center}
\end{figure}

In order to examine the dynamical evolution of the instability of
the KG lattice peakon, we used direct numerical simulations,
performed with a 5th order Calvo's symplectic integrator
\cite{SanzSerna} with a time step $\Delta t=0.001$, which preserves
the energy up to a factor $10^{-15}$. We introduce a perturbation
$\xi_n=\varepsilon\pi_n$
(to the exact peakon solution
$\pi_n$) with $\varepsilon=0.1$ to excite the unstable
eigendirection.
The exponential growth of the peakon instability
is shown in the left panel of Fig.~\ref{fig:simulkg}. If the perturbation were
$\xi_n=-\varepsilon\pi_n$ with $\varepsilon=0.1$ again,
the peakon does not grow. Instead, it evolves to a breathing state
(see e.g. right panel of Fig.~\ref{fig:simulkg}),
which will be analyzed in Section~\ref{sec:breathers}.

\subsection{DNLS 1-site peakons}

As explained above, the (spatial dependence of the) profile of a
DNLS peakon is the same as that of its Klein-Gordon analog (see
Fig.~\ref{fig:peakonkg}). However, in the DNLS setting, the peakon
is stable for all values of $a$. This fact has its origin
in the gauge invariance of the solutions of DNLS-type
equations. In particular, an interesting feature of
the discrete peakons is that the $U(1)$ symmetry of the DNLS chain
prohibits the single negative energy direction that was present
in the KG lattice. Essentially, the unstable direction of the
KG lattice is transversal to the same-charge hypersurface
in the DNLS case. As a result, perturbations along this potentially
unstable direction are {\it banned} by the presence of the extra symmetry.

\begin{figure}
\begin{center}
\begin{tabular}{cc}
\hskip -24pt
    \includegraphics[width=\middlefig]{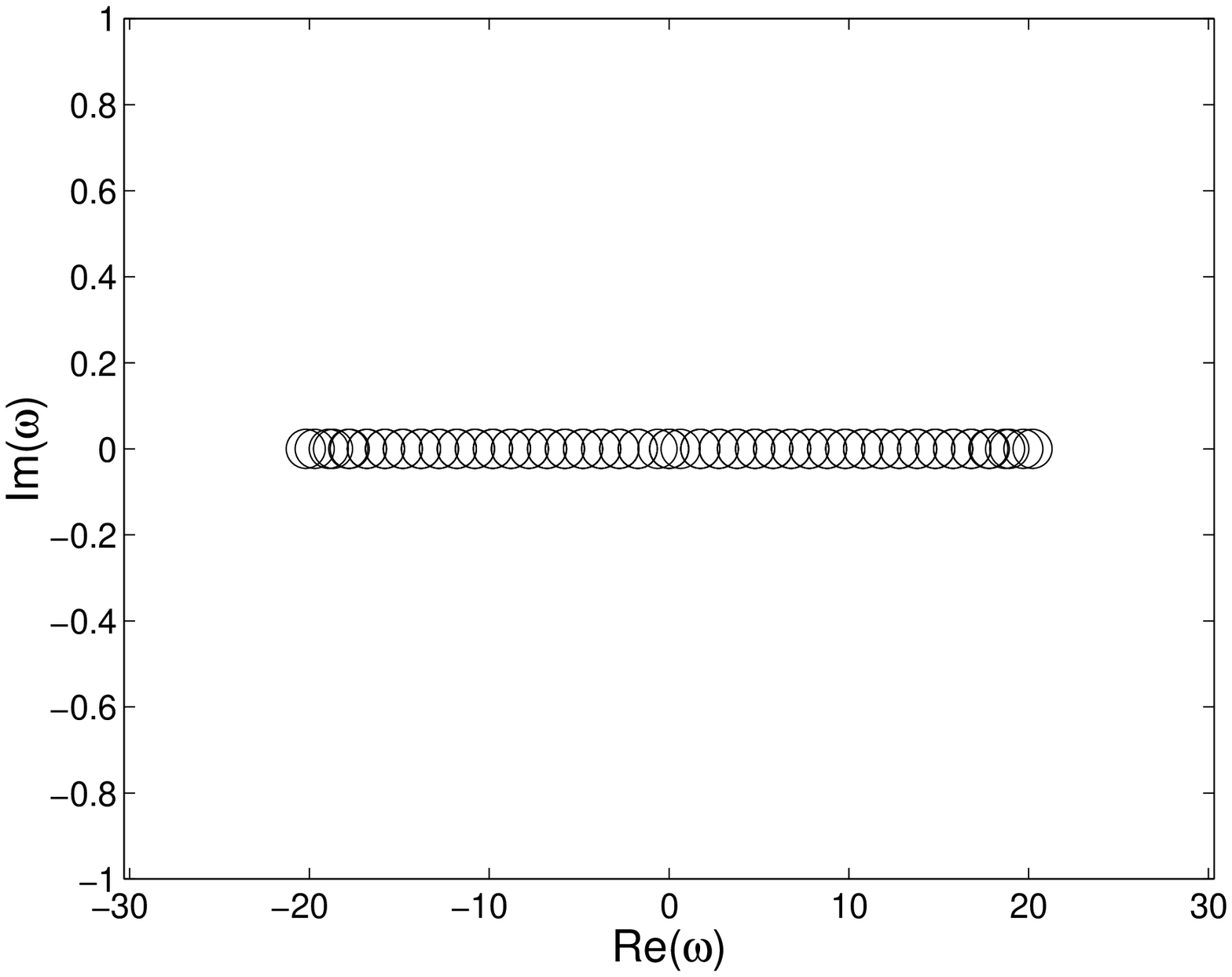} &
    \includegraphics[width=\middlefig]{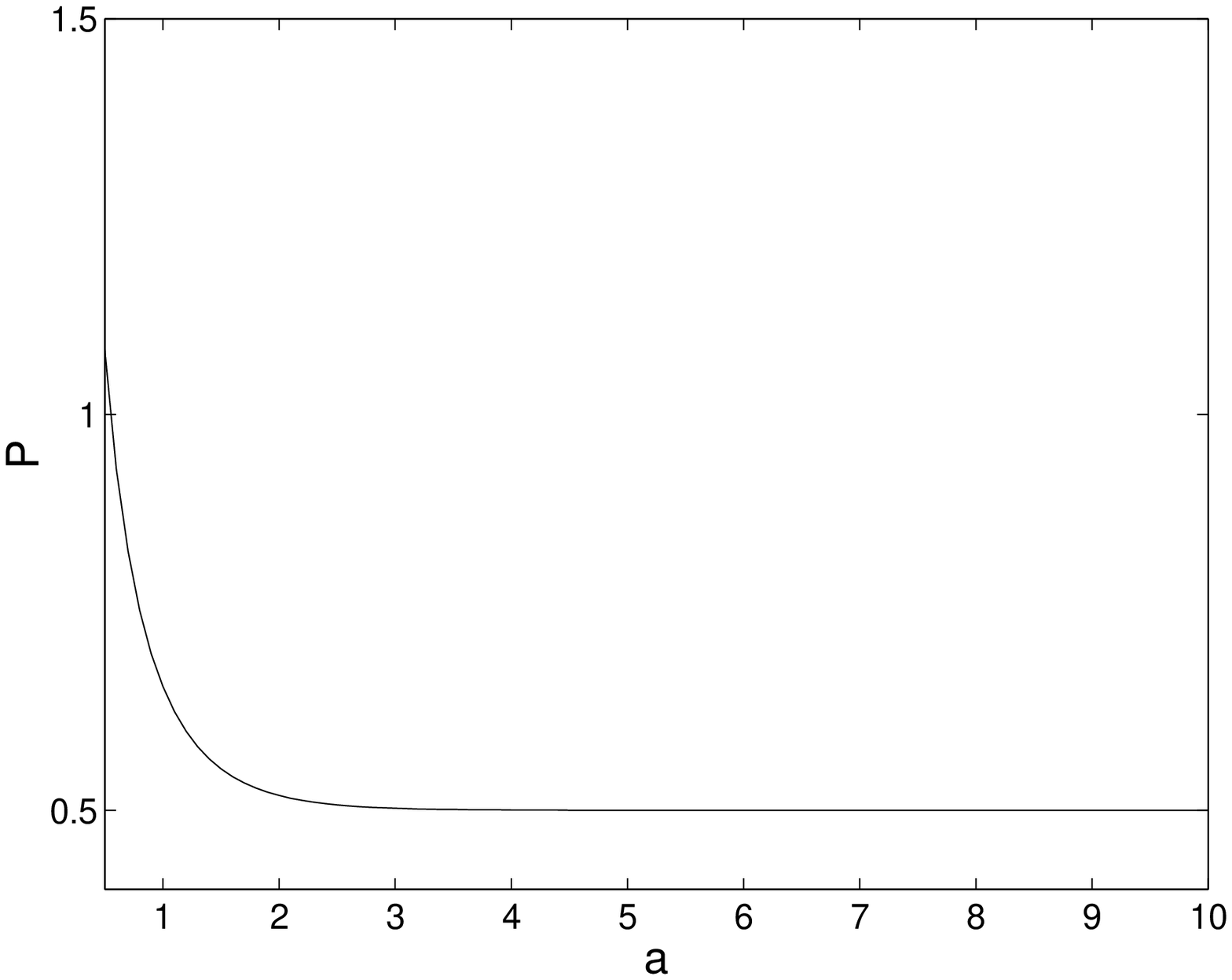}\\
\end{tabular}
\caption{Left panel: Spectral plane of the stability matrix for
the peakon of Fig.~\ref{fig:peakonkg} in the case of the
DNLS chain.
\\
Right panel:
Dependence of the peakon charge (power) $Q=\sum\sb n\pi\sb n^2/2$ as a function of $a$.}
\label{fig:peakondnls}
\end{center}
\end{figure}

Figure~\ref{fig:peakondnls} shows the spectral plane  for a
typical case together with the dependence on $a$ of the charge
(also referred to as power in optics) of the peakon.
The charge is defined as $Q(u)=\sum\sb n|u\sb n|^2/2$
and it can be observed that
its value decreases with $a$ and tends to $Q=1/2$ (as should be
expected as $a \rightarrow \infty$). This dependence can be
analytically calculated: $Q(a)=A^2\coth(a)/2$.

\begin{figure}
\begin{center}
\begin{tabular}{cc}
\hskip -24pt
    \includegraphics[width=\middlefig]{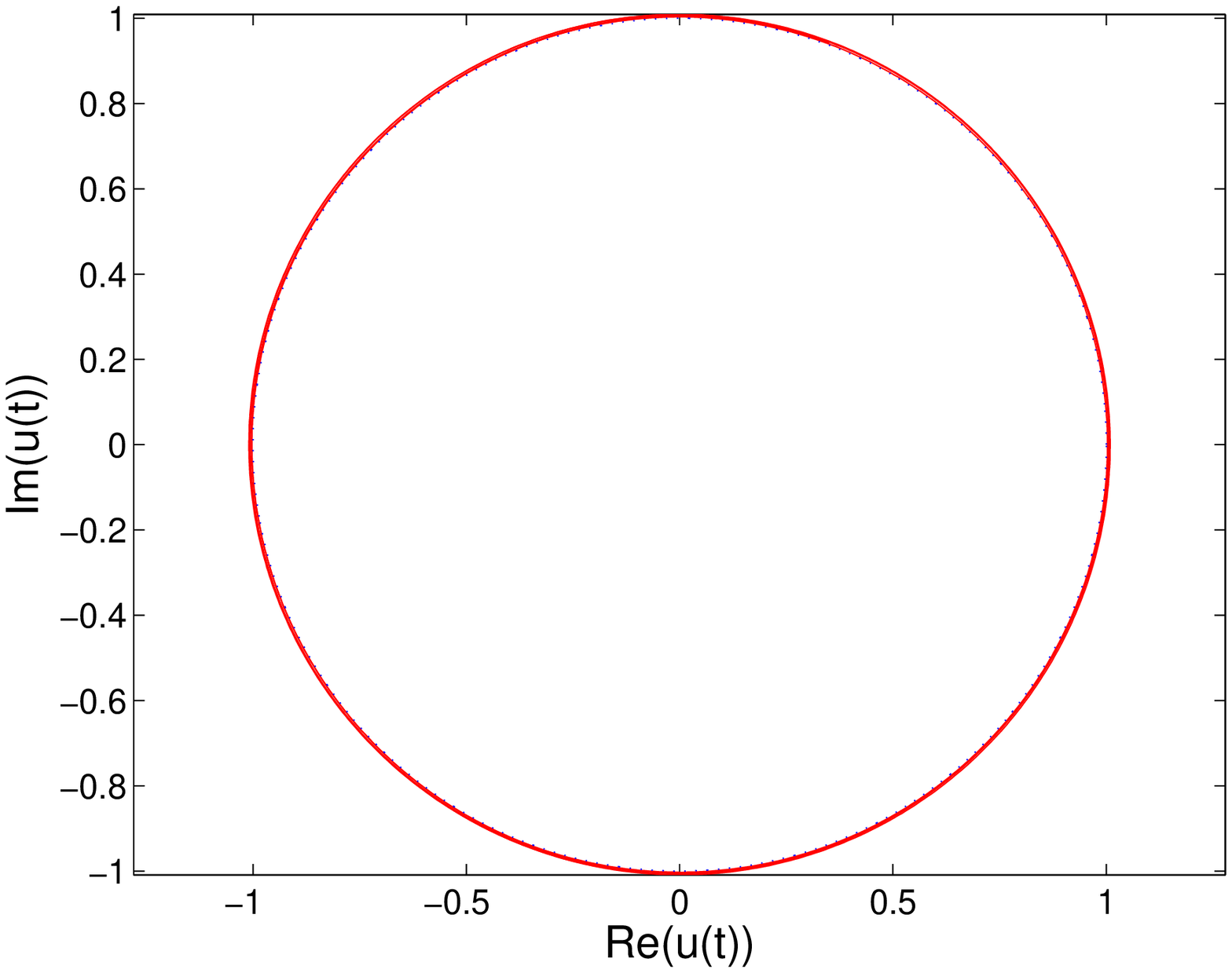} &
    \includegraphics[width=\middlefig]{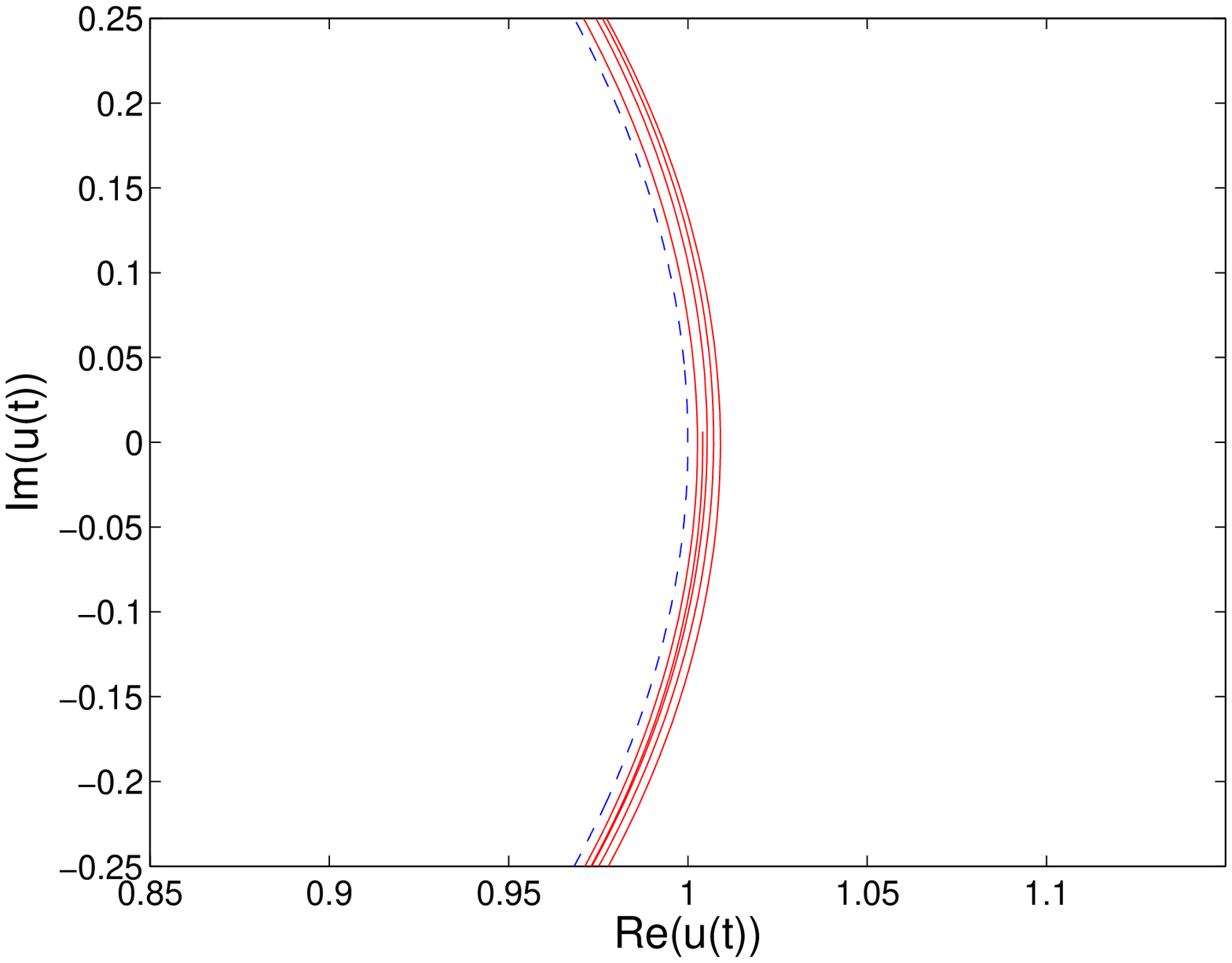}\\
\end{tabular}
\caption{Left panel: Phase space diagram of the central site of a
perturbed (full line) and an unperturbed (dashed line) DNLS peakon.
\\
Right panel: A blow-up of the left panel that illustrates the 
orbital stability of the peakon solution (since the perturbed
solution remains in its vicinity).}
\label{fig:simuldnls}
\end{center}
\end{figure}

The stability of the solution can be verified in the time evolution
numerical experiment shown in Fig.~\ref{fig:simuldnls}, which has
been performed through a 4th order Runge-Kutta integrator with time
step $\Delta t=0.01$. The phase space plot at the central site shows
that a randomly perturbed solution remains orbitally close to the
exact discrete peakon solution.

\subsection{2-site Peakons}

We now examine the behavior of two-site (i.e., inter-site)
peakons, both in the
KG, as well as in the DNLS chain.

The energy and charge of 2-site peakons can be analytically
calculated as $E(a)=B^2/(4a\sinh(a))$ and
$Q(a)=B^2/(2\sinh(a))$.

\begin{figure}
\begin{center}
\begin{tabular}{cc}
\hskip -24pt
    \includegraphics[width=\middlefig]{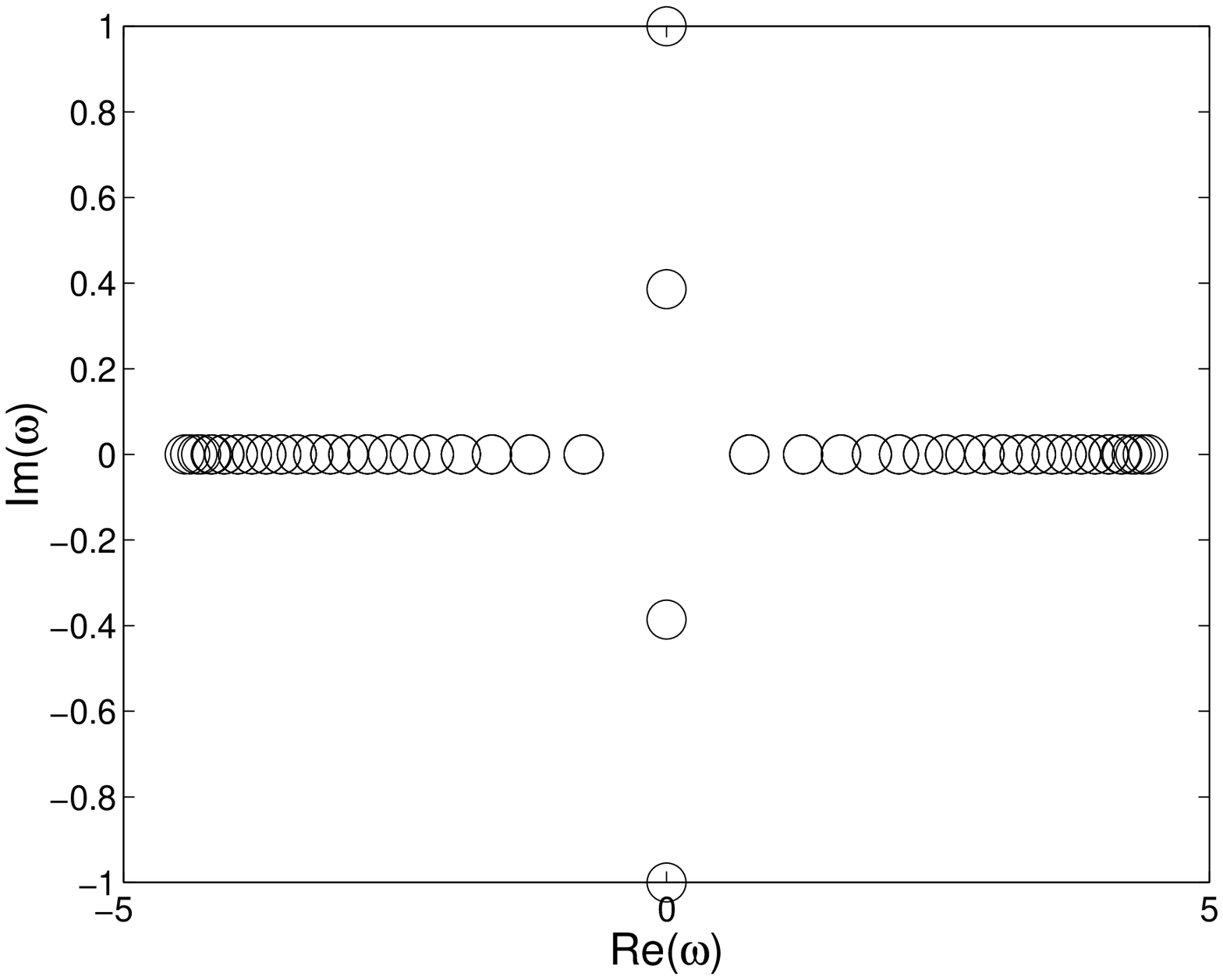} &
    \includegraphics[width=\middlefig]{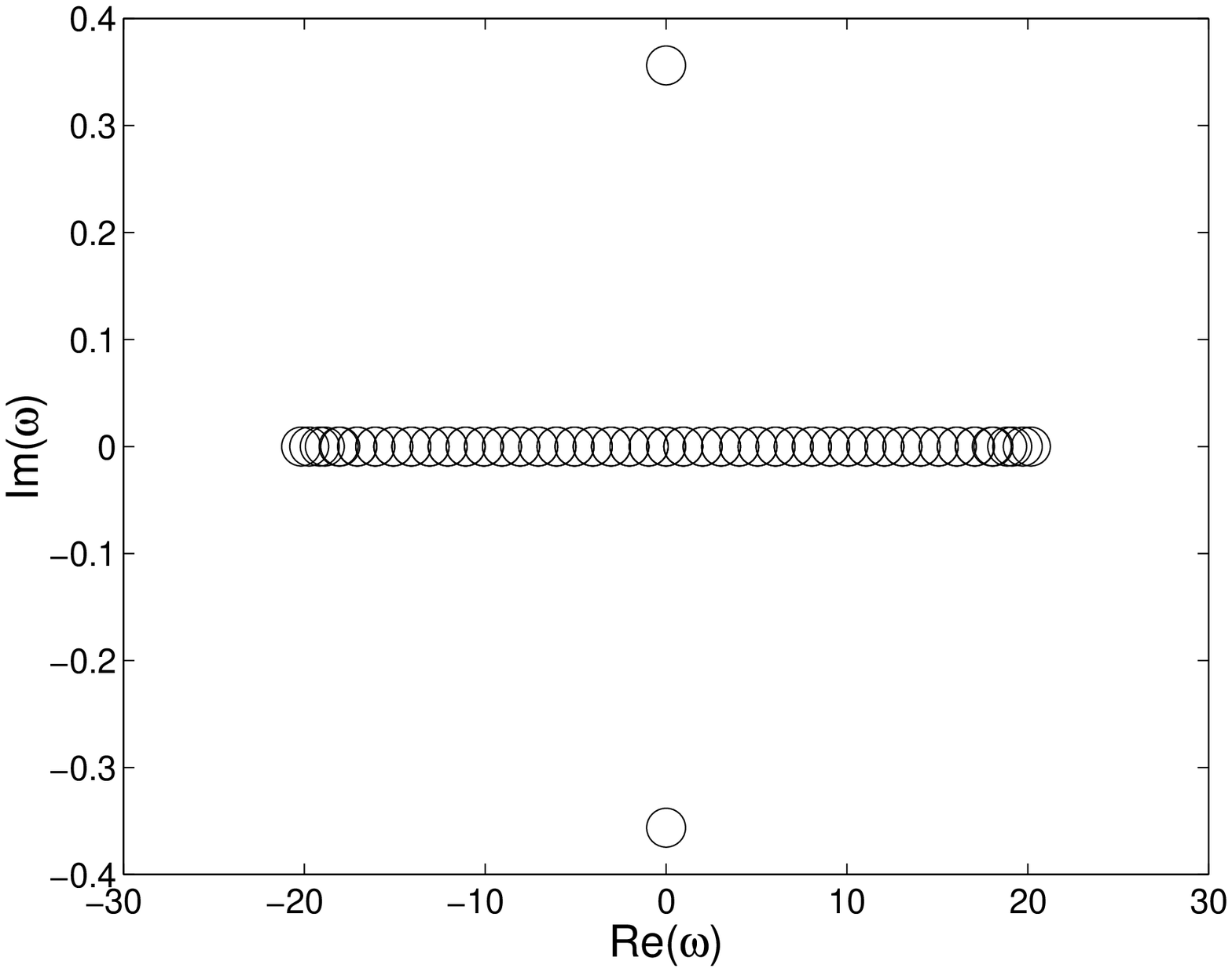}\\
\end{tabular}
\caption{Spectral plane of a the two-site peakon (with $a=1$).
\\
Left panel: KG peakon.
\\
Right panel: DNLS peakon.}%
\label{fig:stab2s}
\end{center}
\end{figure}

Contrary to their single site counterparts, two-site solutions are
unstable (also in the DNLS model).
In this case, two negative energy directions and hence
two imaginary eigenfrequencies could be identified in the spectral
plane of the linearization eigenfrequencies in the case of the KG
lattice, while one such eigenfrequency was present in the DNLS
setting (see Fig.~\ref{fig:stab2s}). The eigenmode corresponding
to the KG case is antisymmetric.

\begin{figure}
\begin{center}
\begin{tabular}{cc}
\hskip -24pt
    \includegraphics[width=\middlefig]{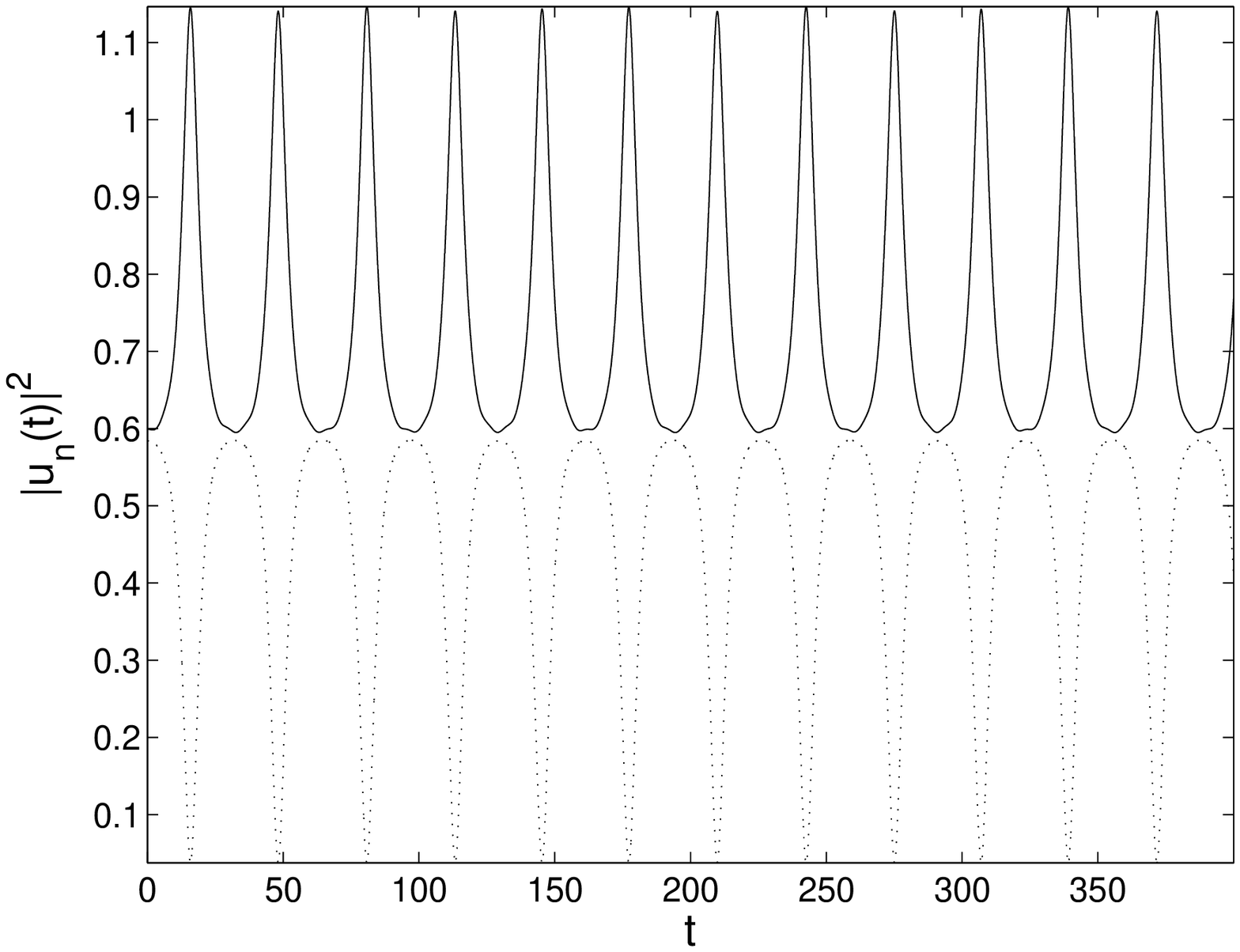} &
    \includegraphics[width=\middlefig]{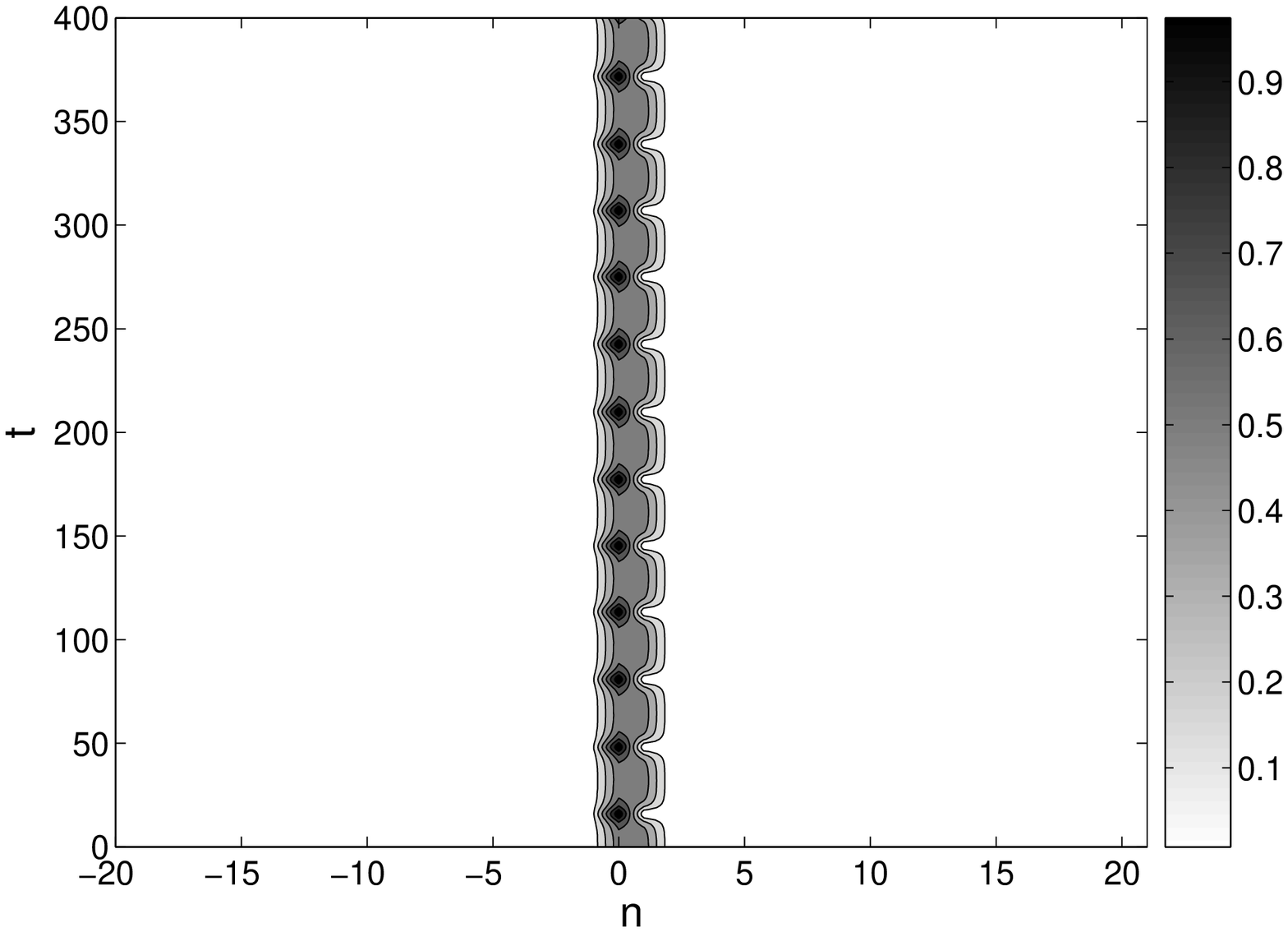}\\
\end{tabular}
\caption{Time evolution of the two-site peakon in the DNLS chain
(with $a=1$) induced by perturbing the central particle.
\\
Left panel:
The full line represents $|u_0|^2$, the dashed line represents $|u_1|^2$.
\\
Right panel: Charge density in space-time.}%
\label{fig:simul2s}
\end{center}
\end{figure}

We also simulated the dynamical development of these
instabilities, observing that KG two-site peakons are completely
destroyed (as their one-site counterparts are also not stable).
For the two-site DNLS peakons excited with the perturbation
$\epsilon\delta\sb{n,0}$ with $\epsilon\sim 10^{-4}$, the solution
oscillates between the one-site and two-site peakons. These
results are shown on Fig.~\ref{fig:simul2s}.

\section{Analytical Results}
\label{sec:analytical}

The above results motivate us to examine the stability of the
Klein-Gordon and DNLS peakons from an analytical perspective
and, in particular, using energetic considerations. We now
proceed to study the stability of the uniform steady state
(``vacuum'') and of one-site peakons
in each of these settings.
The conclusions about stability or instability
do not depend on the values of $A$ and $a$,
so we set
\[
A=1,
\qquad
a=1,
\]
so that the peakon profile is given by
\[
\pi_n=\exp(-|n|).
\]

\subsection{Klein-Gordon}

\subsubsection{Hamiltonian formulation.}

We can rewrite the Klein-Gordon equation (\ref{jeq9}) as
\begin{equation}\label{ddu-is-delta-e-comp}
\ddot u\sb n +\p\sb{u\sb n}T(u)+\p\sb{u\sb n}W(u)=0
\end{equation}
where
\begin{eqnarray}
T(u)=-\frac{1}{2}\sum\sb{(n,m)\in\Z^2}e^{-\abs{n-m}}u\sb n u\sb m,
\label{kg-t}
\\
W(u)=\sum\sb{n\in\Z}
\left(
\left(\frac{\Lambda}{2}+\frac 1 4\right)u\sb n^2
-\frac 1 4 u\sb n^2 \ln u\sb n^2\right).
\label{kg-w}
\end{eqnarray}
Above, $\Lambda$ is a positive constant taken to be
\begin{equation}\label{alpha}
\Lambda
=(\pi,\pi)=\sum\sb{n\in\Z}e^{-2\abs{n}}
=\frac{e^2+1}{e^2-1},
\end{equation}
where
$(u,u)=\sum\sb{n\in\Z}u_n\sp\star u_n$
(in the Klein-Gordon case, we assume that
the components $u\sb n$ are real-valued).

\begin{remark}
Later we will show that the Klein-Gordon equation (\ref{ddu-is-delta-e-comp})
is globally well-posed in $l\sp 2(\Z)\times l\sp 2(\Z)$:
if both $u(0)$ and $\dot u(0)$ belong to $l\sp 2(\Z)$,
then there is a global solution $u(t)$
with $\norm{u(t)}\sb{l\sp 2}<\infty$,
$\norm{\dot u(t)}\sb{l\sp 2}<\infty$
for $0\le t<\infty$.
See Theorem~\ref{theorem-ivp}.
At the same time, the norms
$\norm{u(t)}\sb{l\sp 2}$, $\norm{\dot u(t)}\sb{l\sp 2}$
could grow unboundedly large with time.
\end{remark}

We can rewrite (\ref{ddu-is-delta-e-comp})
as
\begin{equation}\label{ddu-is-delta-e}
\ddot u +T'(u)+W'(u)=0,
\end{equation}
where $T'(u)$, $W'(u)$
may be interpreted as variational derivatives with respect to $u$
of the functionals $T$ and $W$.

The value of the energy functional
\begin{equation}\label{ekg}
E\sb{KG}(u,\dot u)=\sum\sb{n\in\Z}\frac{\dot u\sb n^2}{2}+T(u)+W(u)
\end{equation}
is conserved along the trajectories of (\ref{ddu-is-delta-e}).

\subsubsection{Stability of vacuum.}
First, let us notice that the zero solution is stable
with respect to $l\sp 2$-perturbations of the initial data.
We bound $T(u)$ by
\begin{equation}\label{schur}
\abs{T(u)}
\le
\frac 1 2\Abs{\sup\sb{n}\sum\sb{m\in\Z}e^{-\abs{n-m}}}(u,u)
\le
\frac 1 2\frac{e+1}{e-1}(u,u).
\end{equation}
This inequality is due to the Schur test
applied to the matrix
$J\sb{n m}=e^{-\abs{n-m}}$,
which yields
\begin{equation}\label{schur-test}
\norm{J u}\sb{l\sp 2}
\le
\left(
\sup\sb{n\in\Z}\sum\sb{m\in\Z}\abs{J\sb{nm}}
\right)\sp{\frac 1 2}
\left(
\sup\sb{m\in\Z}\sum\sb{n\in\Z}\abs{J\sb{nm}}
\right)\sp{\frac 1 2}
\norm{u}\sb{l\sp 2}
=
\frac{e+1}{e-1}
\norm{u}\sb{l\sp 2}.
\end{equation}
In the expression
(\ref{kg-w})
for $W(u)$,
the first term is $(\frac{\Lambda}{2}+\frac 1 4)(u,u)$.
When the amplitude of $u$ is small,
the main contribution in (\ref{kg-w})
comes from the last term and is of positive sign;
thus, for
$\norm{u}\sb{l\sp\infty}\equiv\sup\sb{m}\abs{u\sb m}\le\epsilon$,
\begin{equation}
E\sb{KG}(u)
\ge
\left(
-\frac{e+1}{2(e-1)}+\frac{\Lambda}{2}+\frac 1 4
\right)
(u,u)
+
\frac{1}{4}
(u,u)\ln\frac{1}{\epsilon^2},
\end{equation}
which is positive for $\epsilon$ sufficiently small.
Thus, the zero solution minimizes the value of the energy functional
among perturbations with bounded amplitude.
Since
\begin{equation}\label{infty-2}
\norm{u}\sb{l\sp\infty}
=\sup\sb{n\in\Z}\abs{u\sb n}
\le
\left(\sum\sb{n\in\Z}\abs{u\sb n}^2\right)^{1/2}
=\norm{u}\sb{l\sp 2},
\end{equation}
the zero solution also minimizes the energy
among all perturbations with sufficiently small $l\sp 2$-norm.

\subsubsection{Instability of one-site peakons.}
Now we address the stability of the peakon $\pi\sb n=e^{-\abs{n}}$.
Following Derrick \cite{all},
let us consider the family
of vectors $\pi\sp{(\lambda)}$ with the components
$\pi\sb n\sp{(\lambda)}=\lambda\sp{1/2}\pi\sb n$,
where $\pi\sb n$ is the peakon profile.
We have:
\begin{equation}
E\sb{KG}(\pi\sp{(\lambda)})
=T(\pi\sp{(\lambda)})+W(\pi\sp{(\lambda)})
=a\lambda -b\lambda \ln \lambda,
\end{equation}
with $a=T(\pi)+W(\pi)$ and $b=\frac 1 4 \sum\sb{n\in\Z} \pi\sb n^2>0$.

Since $\pi$ is a stationary solution to (\ref{ddu-is-delta-e}),
so that
\begin{equation}\label{stationary-kg}
T'(\pi)+W'(\pi)=0,
\end{equation}
we have:
\begin{equation}
\frac{d}{d\lambda}\At{\lambda=1}
T(\pi\sp{(\lambda)})+W(\pi\sp{(\lambda)})
=(a-b(\ln\lambda+1))\At{\lambda=1}=a-b=0,
\end{equation}
\begin{equation}
\frac{d^2}{d\lambda^2}\At{\lambda=1}
T(\pi\sp{(\lambda)})+W(\pi\sp{(\lambda)})
=-\frac{b}{\lambda}\At{\lambda=1}=-b<0,
\end{equation}
thus the stationary solution $\pi$ does not minimize the energy.

It is also easy to check directly that perturbations
in the direction of the vector
$\frac{d}{d\lambda}\big\vert\sb{\lambda=1}\pi\sp{(\lambda)}
=\frac{1}{2}\pi$
are linearly unstable.
For this, let us
consider the solution of the form $\pi\sb n+r\sb n(t)$;
the linearized equation on $r$ is
\[
\ddot r\sb n+\sum\sb{m\in\Z}(T''(\pi)+W''(\pi))\sb{n m}r\sb m=0,
\]
where
$(T''(\pi)+W''(\pi))\sb{n m}=\frac{\p^2(T+W)}{\p u_n \p u_m}(\pi)$.
The matrix $T''(\pi)+W''(\pi)$ has the eigenvalue $-1$,
with the corresponding eigenvector being $\pi$ itself:
\[
\sum\sb{m\in\Z}(T''(\pi)+W''(\pi))\sb{n m}\pi\sb m=-\pi\sb n.
\]
The perturbation in this direction will grow exponentially, in accordance
with our numerical results (cf. Figs.~\ref{fig:peakonkg}-\ref{fig:simulkg}).

According to \cite{shatah-strauss},
the linear instability in this model
gives rise to the dynamic, or {\it nonlinear},
instability:

\begin{theorem}
The stationary peakon solution $u(t)=\pi$
to {\rm (\ref{ddu-is-delta-e})}
is unstable with respect to $l\sp 2$-perturbations
of the initial data.
\end{theorem}

\subsection{DNLS}

\subsubsection{Hamiltonian formulation.}

We now turn to the DNLS equation (\ref{jeq10}), which can be rewritten as:
\begin{equation}\label{du-is-delta-e}
i\dot u_n=2\p\sb{u\sp\star_n}E(u,u\sp\star),
\end{equation}
where
$E(u,u\sp\star)=T(u,u\sp\star)+V(u,u\sp\star)$, with
\begin{eqnarray}
T(u,u\sp\star)=-\frac{1}{2}\sum\sb{(n,m)\in\Z^2}
e^{-\abs{n-m}}u\sp\star\sb n u\sb m,
\label{nls-t}
\\
V(u,u\sp\star)
=\sum\sb{n\in\Z}
\left(
\left(\frac{\Lambda}{2}-\frac 1 4\right)\abs{u\sb n}^2
-\frac 1 4 \abs{u\sb n}^2 \ln \abs{u\sb n}^2\right).
\label{nls-g}
\end{eqnarray}
The value of the energy functional
\[
E(u,u\sp\star)=T(u,u\sp\star)+V(u,u\sp\star)
\]
is conserved along the trajectories of (\ref{du-is-delta-e}).
The value of the charge functional
\[
Q(u,u\sp\star)
=\frac 1 2\sum\sb{n\in\Z} u\sb n u\sp\star\sb n
=\frac 1 2(u,u)
\]
is also conserved (due to the $U(1)$-invariance of the system).

\subsubsection{Stability of vacuum.}

Stability of the vacuum solution $u=0$
with respect to $l\sp 2$-perturbations
of the initial data
is proved in the same way as in the case of Klein-Gordon equation:
The solution $u=0$
delivers the smallest (zero) value to the energy functional
among all vectors $\psi$
of sufficiently small $l\sp 2$-norm.

\subsubsection{Stability of one-site DNLS peakons.}

In the case of DNLS, the perturbation
$u\mapsto u+\epsilon u$
would lead to smaller values of the energy functional
as in the case of Klein-Gordon equation.
However, now this perturbation
is prohibited  by the charge conservation.
Therefore, we expect that the peakons are local
minimizers of the energy functional under the charge constraint and hence are
orbitally stable.

We will prove the following theorem:

\begin{theorem}\label{theorem-dnls-stability}
The standing wave solution
$e^{it}\pi$
to {\rm (\ref{jeq10})}
is orbitally stable
with respect to $l\sp 2$-perturbations
of the initial data.
\end{theorem}

Let us remind the definition of the orbital stability
(see \cite{strauss}):

\begin{definition}
\label{definition-stability}
The $\pi$-orbit
$\{ e^{i s}\pi\sothat s\in\R\}$
is {\it stable}
if for all $\varepsilon>0$
there exists $\delta>0$
with the following property.
If $u\sb 0\in l\sp 2(\Z,\C)$
is such that
$\norm{u\sb 0-\pi}\sb{l\sp 2}<\delta$
and $u(t)$ is a solution of Eq.~(\ref{jeq10})
in some interval $[0,t\sb 1)$,
then $u(t)$ can be continued to a solution in $0\le t<\infty$
and
\[
\sup\sb{0<t<\infty}
\inf\sb{s\in\R}
\norm{u(t)-e^{i s}\pi}\sb{l\sp 2}<\varepsilon.
\]
Otherwise the $\pi$-orbit is called {\it unstable}.
\end{definition}

\begin{remark}
Eq.~(\ref{jeq10})
is locally well-posed in $l\sp 2(\Z,\C)$
since for $u\in l\sp 2$ the right-hand side
of (\ref{jeq10})
(where we set $A=1$, $a=1$)
is also in $l\sp 2$:
\begin{equation}\label{also-l2}
\Norm{-\sum\sb{m\in\Z}J\sb{n m} u\sb m
+\left[
\frac{2}{e^2-1}
-\frac{1}{2}\ln\left(
u\sb n u\sb n\sp\star
\right)
\right]u\sb n
 }\sb{l\sp 2}
\le
\left[
\frac{e+1}{e-1}
+
\frac{2}{e^2-1}
+\Abs{\ln
\norm{u}\sb{l\sp 2}
}
\right]
\norm{u}\sb{l\sp 2}.
\end{equation}
We used inequalities (\ref{schur-test}) and (\ref{infty-2}).
Hence, for any $u\sb 0\in l\sp 2(\Z,\C)$,
there exists $\tau>0$ so that there is a unique
solution $u(t)$ defined for $0\le t<\tau$
that satisfies $u(0)=u\sb 0$.
Due to the conservation
of the charge $Q(u)=\norm{u}\sb{l\sp 2}^2/2$
along the flow of Eq.~(\ref{jeq10}),
we conclude that $u(t)$ is defined globally:
$u\in C\sp 1([0,\infty),l\sp 2(\Z,\C))$.
This settles the question about the
global well-posedness for (\ref{jeq10}) in $l\sp 2(\Z,\C)$
which is a necessary condition for the orbital stability.
\end{remark}

According to \cite{strauss},
we will know the orbital stability of the peakon if we can prove that
$
H=E''(\pi)+Q''(\pi)
$
defines a quadratic form that is positive-definite on
vectors that are tangent to the hypersurface of the same
charge and are orthogonal to the orbit spanned by $\pi$.

\begin{remark}\label{remark-rest}
Once one knows that $H$ defines a positive-definite quadratic
form on vectors tangent to same charge hypersurface and orthogonal
to the orbit of $\pi$, one proves that
there exist $\delta>0$ and $C>0$
such that for any $u$ with $Q(u)=Q(\pi)$
and $\norm{u-\pi}\sb{l\sp 2}<\delta$
one has $E(u)-E(\pi)\ge C\inf\sb{s}\norm{u-e^{i s}\pi}\sb{l\sp 2}$
(Theorem 3.4 in \cite{strauss})
and then the orbital stability follows from Theorem 3.5 in \cite{strauss}.
\end{remark}

We will use the real-valued formulation.
For $u(t)=v(t)+iw(t)\in l\sp 2(\Z,\C)$,
with $v=\{v\sb n\}$, $w=\{w\sb n\}$
real-valued,
we write
\[
\bm u(t)=\left[\begin{array}{c}v(t)\\ w(t)\end{array}\right],
\qquad
\bm u(t)\in l\sp 2(\Z,\R^2).
\]
The equation on $\bm u$ is
\[
\dot{\bm u}=J E'(\bm u)
=
\left[\begin{array}{cc}0&1\\-1&0\end{array}\right]
\left[\begin{array}{c}\nabla\sb v E\\\nabla\sb w E\end{array}\right],
\]
and the stationary equation on
$
\bm\pi=\left[\begin{array}{c}\pi\\ 0\end{array}\right]
$
is given by
\[
E'(\bm\pi)+Q'(\bm\pi)=0.
\]

Let
$
\psi=\xi+i\eta,
$
$
\bm\psi=\left[\begin{array}{c}\xi\\\eta\end{array}\right].
$
The vectors tangent to the same charge hypersurface satisfy
\begin{equation}\label{same-charge}
\langle Q'(\bm\pi),\bm\psi\rangle
=\langle \bm\pi,\bm\psi\rangle
=\sum\sb{n\in\Z}\pi\sb n\xi\sb n=0,
\end{equation}
while the vectors orthogonal to
$\left[\begin{array}{c}0\\\pi\end{array}\right]=-J\bm\pi$
(this vector corresponds to $i \pi$,
a tangent direction to the orbit spanned by $e^{i t}\pi$)
satisfy
\begin{equation}\label{orthogonal-to-orbit}
\langle -J\bm\pi,\bm\psi\rangle
=-\sum\sb{n\in\Z}\pi\sb n\eta\sb n=0.
\end{equation}

As we mentioned in Remark~\ref{remark-rest},
Theorem~\ref{theorem-dnls-stability}
will be proved
if we can show that the quadratic form
defined by the Hamiltonian operator
\[
H=E''(\bm\pi)+Q''(\bm\pi)
\]
is positive-definite on vectors
$\bm\psi=\left[\begin{array}{c}\xi\\\eta\end{array}\right]$,
where both $\xi$ and $\eta$ are orthogonal to $\pi$.

Let us find the explicit expression for $H$.
For the second Fr\'echet derivative of $T$,
we compute:
\begin{equation}
\langle T''(\bm\pi)\bm\psi,\bm\psi\rangle
=-\sum\sb{(n,m)\in\Z^2}
e^{-\abs{n-m}}
\left(\xi\sb n\xi\sb m+\eta\sb n\eta\sb m\right).
\end{equation}
For the ``potential'' term, we compute
\begin{equation}
\langle
V''(\bm\pi)\bm\psi,\bm\psi\rangle
=\sum\sb{n\in\Z}\left(
(\Lambda-1)(\xi^2\sb n+\eta^2\sb n)
+
\abs{n}(\xi^2\sb n+\eta^2\sb n)-\xi^2\sb n
\right).
\end{equation}
For the charge functional, we have
\begin{equation}
\langle Q''(\bm\pi)\bm\psi,\bm\psi\rangle
=\sum\sb{n\in\Z}\left(
\xi^2\sb n+\eta^2\sb n
\right).
\end{equation}
Hence, the operator
$H=E''(\bm\pi)+Q''(\bm\pi)=T''(\bm\pi)+V''(\bm\pi)+Q''(\bm\pi)$
is given by
\begin{eqnarray}
\langle H\bm\psi,\bm\psi\rangle
&=
-\sum\sb{(n,m)\in\Z^2} e^{-\abs{n-m}}
\left(\xi\sb n\xi\sb m+\eta\sb n\eta\sb m\right)
\\
&\quad\qquad
+\sum\sb{n\in\Z}\left(
\left(\Lambda-1+\abs{n}\right)
(\xi^2\sb n+\eta^2\sb n)+\eta^2\sb n
\right).
\nonumber
\end{eqnarray}

We have:
\begin{equation}\label{h-pm}
\langle H\bm\psi,\bm\psi\rangle
=
\left[\xi,\eta\right]
\left[\begin{array}{cc}H\sp{+}&0\\0&H\sp{-}\end{array}\right]
\left[\begin{array}{c}\xi\\ \eta\end{array}\right],
\end{equation}
where
\begin{eqnarray}
H\sp{+}\sb{n m}
=-e^{-\abs{n-m}}+\left(\Lambda-1+\abs{n}\right)\delta\sb{n m}
\label{q1-better-be-positive}
\\
H\sp{-}\sb{n m}
=-e^{-\abs{n-m}}+\left(\Lambda+\abs{n}\right)\delta\sb{n m}.
\label{q2-better-be-positive}
\end{eqnarray}
Let us consider the symmetric matrix
\begin{equation}\label{matrix-a}
A\sb{n m}
=-e^{-\abs{n-m}}
+\abs{n}\delta\sb{n m}
+\pi\sb n \pi\sb m.
\end{equation}
Numerical computations show that this matrix does not have negative
eigenvalues and
that the dimension of its null space is $\dim N(A)=3$.
One can check that
$N(A)$ is spanned by the vectors
$v\sp{(+)}$, $v\sp{(-)}$, and $v\sp{(0)}$:
\[
v\sp{(+)}\sb n=\theta(n+1/2)e^{-\abs{n}},
\]
\[
v\sp{(-)}\sb n
=v\sp{(+)}\sb{-n}
=\theta(-n+1/2)e^{-\abs{n}},
\]
\[
v\sp{(0)}\sb n=\delta\sb{n,0}
\]
where $\theta(n)=1$ for $n>0$
and $0$ for $n<0$.
In particular,
\[
\pi
=v\sp{(+)}+v\sp{(-)}-v\sp{(0)}\in N(A).
\]
Since $A$ is symmetric,
the eigenvectors that correspond to other (positive)
eigenvalues of $A$ are orthogonal to $\pi$.

The matrices $H\sb{n m}\sp{+}$, $H\sb{n m}\sp{-}$
can be expressed as
\begin{eqnarray}
H\sb{n m}\sp{+}=A\sb{n m}-\pi\sb n \pi\sb m+(\Lambda-1)\delta\sb{n m},
\label{h-plus}
\\
H\sb{n m}\sp{-}=A\sb{n m}-\pi\sb n \pi\sb m+\Lambda\delta\sb{n m}
=H\sb{n m}\sp{+}+\delta\sb{n m}.
\label{h-minus}
\end{eqnarray}

We have:
$H\sp{+}\pi=A\pi-(\pi,\pi)\pi+(\Lambda-1)\pi=-\pi$,
since $(\pi,\pi)=\Lambda$,
and $H\sp{-}\pi=0$.
Since $H\sp{+}$ and $H\sp{-}$ are symmetric,
their other eigenvectors are orthogonal to $\pi$.
But then
they also have to be the eigenvectors of $A$.
The corresponding eigenvalues are those of $A$
shifted to the right by $(\pi,\pi)-1>0$ (for $H\sp{+}$)
and by $(\pi,\pi)$ (for $H\sp{-}$),
and hence are strictly positive.

\begin{remark}
The important feature of this model that makes the analysis simple
is that $H\sp{+}$ and $H\sp{-}$
can be diagonalized simultaneously.
\end{remark}

Thus, assuming that $\psi=\xi+i\eta\in l\sp 2(\Z,\C)$
satisfies
(\ref{same-charge}) and (\ref{orthogonal-to-orbit})
(both $\xi$ and $\eta$ are orthogonal to $\pi$)
and is different from zero,
we conclude that $\langle H\bm\psi,\bm\psi\rangle$ is positive-definite:
\[
\langle H\bm\psi,\bm\psi\rangle
=
\sum\sb{(n,m)\in\Z^2}
\left(
H\sp{+}\sb{nm}\xi\sb n\xi\sb m
+
H\sp{-}\sb{nm}\eta\sb n\eta\sb m
\right)
>0
\]
and hence the one-site peakon solutions are dynamically stable.

\subsubsection{Two-site DNLS peakons}

One can also prove the following theorem:
\begin{theorem}\label{theorem-dnls-stability-2}
The 2-site standing wave solution to {\rm (\ref{jeq10})}
is orbitally unstable with respect to $l\sp 2$-perturbations of the
initial data.
\end{theorem}

The generalization of the matrices $H\sp{+}$ and $H\sp{-}$ given
above in Eq.~(\ref{q1-better-be-positive}) and
(\ref{q2-better-be-positive}) is given by:
\begin{eqnarray}
H\sp{+}\sb{n m}
=-e^{-\abs{n-m}}+\left(\Lambda-1-\ln (u_n)\right)\delta\sb{n m}
\label{q1-better-be-positive-again}
\\
H\sp{-}\sb{n m}
=-e^{-\abs{n-m}}+\left(\Lambda+ \ln(u_n) \right)\delta\sb{n m}.
\label{q2-better-be-positive-again}
\end{eqnarray}

In the case of the inter-site solution, known explicitly as
\begin{eqnarray}
u_n=\frac{1}{2} \frac{(e-1)}{e+1} e^{i t} e^{-|n-1/2|},
\label{2-site-steady}
\end{eqnarray}
\noindent
\hskip -20pt

\noindent
the eigenvalues of $H\sp{+}$, $H\sp{-}$ can be explicitly computed. As
expected, $u_n$ is itself an eigenvector with eigenvalue $-1$ (see above),
while the second eigenvalue is $\lambda_2=-0.149$. However,
since the eigenvalues of $H\sp{-}$ are the ones of $H\sp{+}$
shifted by $1$, this results in $n(H\sp{+})-n(H\sp{-})=2$, where
$n(H\sp\pm)$ denotes the number of negative eigenvalues
of the matrices $H\sp\pm$. From the theory of \cite{strauss,jones,grillakis}
(see also the recent work of \cite{kapitula,pelinovsky}), this implies
that there is one real eigenvalue in the spectrum of linearization
of the DNLS 2-site peakon, in agreement with our numerical observations
of Section~\ref{sec:numerical}.
This completes the proof of Theorem~\ref{theorem-dnls-stability-2}.

\section{Breathing peakons in the discrete Klein-Gordon equation}
\label{sec:breathers}

\subsection{Existence of Breathing Peakons}

We are interested in solutions
to the Klein-Gordon equation (\ref{ddu-is-delta-e-comp})
that have the form
\begin{equation}\label{breathers}
u\sb n(t)=g(t)f\sb n,
\end{equation}
where $g(t)$ is a scalar-valued function.

\begin{figure}[htbp]
\begin{center}

\includegraphics[width=\middlefig]{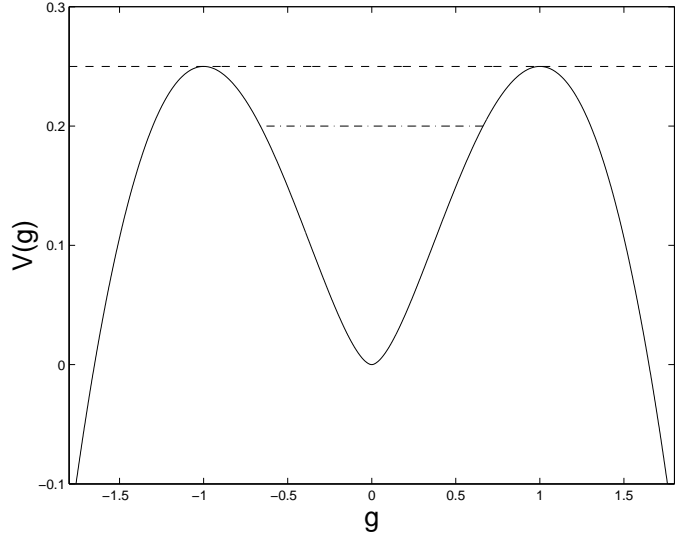}
\caption{ Potential ${\mathcal V}(g)$. The (unstable) stationary
solutions $g=\pm 1$ correspond to the (unstable) peakons, $u(t)
=g(t)\pi=\pm\pi$, while the (stable) stationary solution $g=0$
corresponds to the vacuum (which we know to be stable). There are
also oscillating solutions that correspond to breathing
oscillations with the energy ${\mathcal E}$, $0\le {\mathcal E}<1/4$. A
breathing ``bound state'' at ${\mathcal E}=0.2$ is also shown by
dash-dotted line.}
\label{breathers-g-potential}
\end{center}
\end{figure}

Trying the peakon profile, $f=\pi$,
so that $u\sb n(t)=g(t)e^{-\abs{n}}$,
we get the following equation on $g(t)$:
\begin{equation}\label{eqn-on-g}
\ddot g=\frac 1 2 g\ln g^2, \qquad g=g(t).
\end{equation}
We can integrate Eq.~(\ref{eqn-on-g}), getting
\begin{equation}\label{eqn-on-g2}
\frac{\dot g^2}{2}+\frac{g^2}{4}(1-\ln g^2)={\mathcal E},
\end{equation}
where ${\mathcal E}\ge 0$ is the ``energy of breathing''.
The function ${\mathcal V}(g)=\frac{g^2}{4}(1-\ln g^2)$
(see Fig.~\ref{breathers-g-potential})
represents the potential
in which $g$ lives.

\subsection{
Global well-posedness for the discrete Klein-Gordon equation in $l\sp 2\times l\sp 2$}

Before analyzing the stability of the breathing peakons,
we need to consider the initial value problem
for the equation of the Klein-Gordon lattice
(\ref{ddu-is-delta-e-comp}).

\begin{theorem}\label{theorem-ivp}
Eq.~(\ref{ddu-is-delta-e-comp})
is globally well-posed in $l\sp 2\times l\sp 2$.
That is, for any initial data
$(u\sb 0,v\sb 0)\in l\sp 2(\Z)\times l\sp 2(\Z)$,
there is a unique solution $u(t)$
that satisfies $u(0)=u\sb 0$, $\dot u(0)=v\sb 0$;
this solution is defined for all times $t\ge 0$,
and moreover
$\norm{u(t)}\sb{l\sp 2}+\norm{\dot u(t)}\sb{l\sp 2}$
remains finite for all $t\ge 0$:
\[
u\in C\sp 2([0,\infty),l\sp 2(\Z)).
\]
\end{theorem}

Let us first discuss
the local well-posedness of (\ref{ddu-is-delta-e-comp})
in $l\sp 2$.
We claim that
for any $(u\sb 0,v\sb 0)\in l\sp 2(\Z)\times l\sp 2(\Z)$,
there exists $\tau>0$
such that there is a unique solution
$u(t)$
defined for $0\le t<\tau$
with $(u,\dot u)=(u\sb 0,v\sb 0)$
and $u\in C^1([0,\tau),l\sp 2(\Z))$.
This claim immediately follows from the observation
that if $u\in l\sp 2(\Z)$,
then $\ddot u$ given by (\ref{ddu-is-delta-e-comp})
is also from $l\sp 2(\Z)$.
This is verified as in (\ref{also-l2}).

Now we turn to the global well-posedness.
Let us prove the boundedness of
$L(t)=\norm{u(t)}\sb{l\sp 2}^2$.
Let us derive the equation for $d^2 L(t)/dt^2$.
We have:
\[
\frac{d^2}{dt^2}L(t)
=2(\dot u,\dot u)+2(u,\ddot u).
\]
The total energy
$
\frac{1}{2}(\dot u,\dot u)+T(u)+W(u)
$
is conserved along the flow generated by
(\ref{ddu-is-delta-e-comp})
allowing to express
\begin{equation}\label{dot-u-squared}
(\dot u,\dot u)=2 E\sb 0-2T(u)-2W(u),
\end{equation}
where $E\sb 0=(v\sb 0,v\sb 0)/2+T(u\sb 0)+W(u\sb 0)$
and $(u\sb 0,v\sb 0)$ is the initial data
that corresponds to the solution $u(t)$.

Thus, we obtain:
\begin{eqnarray}
\frac{d^2}{dt^2}L(t)
&=&4E\sb 0-4T(u)-4W(u)-2(u,\p\sb u T)-2(u,\p\sb u W)
\nonumber
\\
&=&4E\sb 0-8T(u)
+\sum\sb{n\in\Z}
\left(
-\left(4\Lambda+1\right)u\sb n^2+2 u\sb n^2\ln u\sb n^2
\right).
\nonumber
\end{eqnarray}
Using the straightforward bounds on
the terms in the right-hand side,
namely
\begin{equation}\label{bound-on-t}
-T(u)\le \frac{e+1}{2(e-1)}L
\end{equation}
that follows from (\ref{schur}),
and also
\begin{equation}\label{bound-on-log}
\sum\sb{n\in\Z}u\sb n^2\ln u\sb n^2
\le\sum\sb{n\in\Z}u\sb n^2\ln \norm{u}\sb{l\sp\infty}^2
\le\sum\sb{n\in\Z} u\sb n^2\ln L
= L\ln L,
\end{equation}
where in the second inequality we used relation (\ref{infty-2}),
we conclude that, for some $C=C(E\sb 0)>0$,
\begin{equation}
\frac{d^2}{dt^2}L
\le C(1+L+2 L\ln L).
\label{ddot-q}
\end{equation}
For $L\ge 1$,
the right-hand side is monotonically increasing
(and exceeds its range for $0<L<1$).
Therefore, $0\le L(t)\le Z(t)$,
where $Z(t)$ is a function that satisfies
\begin{equation}\label{zpp}
Z''(t)=C(1+Z+2 Z\ln Z)
\end{equation}
and the initial data
\[
Z(0)=\max(1,L(0))=\max(1,(u\sb 0,u\sb 0)),
\]
\[
Z'(0)
=\max\left(1,\frac{dL}{dt}\At{t=0}\right)
=\max\left(1,2(u\sb 0,v\sb 0)\right).
\]
We can rewrite (\ref{zpp}) as
\begin{equation}
Z''+C\p\sb Z(-Z-Z^2\ln Z)=0;
\end{equation}
multiplying by $Z'$ and integrating in $t$, we get
\[
\frac{(Z')^2}{2}-C Z-C Z^2\ln Z=E,
\]
where
$E=\frac{(Z'(0))^2}{2}+C(-Z(0)-Z^2(0)\ln Z(0))
$
is a constant of integration.
Expressing $Z'$ and separating variables, we get
\[
t=
\int\sb{Z(0)}\sp{Z(t)}
\frac{dZ}{\sqrt{E+C Z+C Z^2\ln Z}}.
\]
Since the integral
$
\int\sb{Z(0)}\sp\infty
\frac{dZ}{\sqrt{E+C Z+C Z^2\ln Z}}$
diverges at the upper limit
(as $\ln\sp{\frac 1 2} Z$),
we conclude that
$Z$ can not become infinite in finite time.

The finiteness of $l\sp 2$-norm of $\dot u$
follows from relation (\ref{dot-u-squared}),
bounds (\ref{bound-on-t}) and (\ref{bound-on-log}),
and the finiteness of $L=\norm{u}\sb{l\sp 2}^2$
that we already proved.

This finishes the proof of Theorem~\ref{theorem-ivp}.

\subsection{Linearized stability of Breathing Peakons}

We will now analyze the linear stability
of the breathing peakon, showing the
absence of the exponential instability
of small perturbations of the initial data.
We rewrite Eq.~(\ref{ddu-is-delta-e-comp})
as the first order system,
\begin{equation}\label{kg-3}
\left\{
\begin{array}{l}
\dot u=v
\\
\dot v=-\p\sb u(T(u)+W(u)),
\end{array}
\right.
\end{equation}
and consider the perturbation of the solution
$(u\sb 0(t),v\sb 0(t))=(g(t)\pi,\dot g(t)\pi)$
that corresponds to the peakon:
\[
u(t)=u\sb 0(t+\gamma(t))+\delta u(t),
\qquad
v(t)=v\sb 0(t+\gamma(t))+\delta v(t).
\]
The function $\gamma(t)$ adjusts the location of
the breather $(g(t)\pi,\dot g(t)\pi)$
so that it is closer (in a certain sense)
to the perturbed solution $(u(t),v(t))$.
We consider the linearization of system (\ref{kg-3}).
For this, we first compute
\begin{equation}\label{ekgpp}
T''(g\pi)+W''(g\pi)
=T''(\pi)+W''(\pi)-\frac{\ln g^2}{2}
=H\sp{+}-\frac{\ln g^2}{2},
\end{equation}
where $H\sp{+}$ was introduced in (\ref{h-plus}),
and then we can write
\begin{equation}\label{kg-breather-linearized}
\left\{
\begin{array}{l}
g'(t+\gamma(t))\dot\gamma(t)\pi
+\p\sb t\delta u(t)=\delta v(t),
\\
g''(t+\gamma(t))\dot\gamma(t)\pi
+\p\sb t\delta v(t)=-(H\sp{+}-\frac{\ln g^2}{2})\delta u(t).
\end{array}
\right.
\end{equation}
We split
\[
\delta u(t)=a(t)\pi+\phi(t),
\qquad
\delta v(t)=b(t)\pi+\psi(t),
\]
with $\phi$, $\psi\in l\sp 2(\Z)$
both orthogonal to $\pi$.
Projecting system (\ref{kg-breather-linearized})
onto $\pi$, gives the following system:
\begin{equation}\label{kg-breather-linearized-pi}
\left\{
\begin{array}{l}
g'(t+\gamma(t))\dot\gamma(t)
+\dot a(t)=b(t),
\\
g''(t+\gamma(t))\dot\gamma(t)\pi
+\dot b(t)=-(-1-\frac{\ln g^2}{2})a(t),
\end{array}
\right.
\end{equation}
where we used the fact that $\pi$ is an eigenvector,
$H\sp{+}\pi=-\pi$.
Projection of (\ref{kg-breather-linearized})
onto the direction normal to $\pi$ gives the following system:
\begin{equation}\label{kg-breather-linearized-nopi}
\left\{
\begin{array}{l}
\dot\phi(t)=\psi(t),
\\
\dot\psi(t)=-(H\sp{+}-\frac{\ln g^2}{2})\phi(t).
\end{array}
\right.
\end{equation}
The analysis of system (\ref{kg-breather-linearized-pi})
is straightforward.
We would arrive at this system
if we pursued the stability analysis
of Eq.~(\ref{eqn-on-g}).
At the same time, we could analyze that equation topologically.
It corresponds to an unharmonic oscillator;
its phase portrait in the $(g,\dot g)$ plane
contains a set of closed trajectories circling around the origin
(they correspond to the initial data $(g,\dot g)$
such that $\abs{g}<1$ and the value of ${\mathcal E}$
in (\ref{eqn-on-g2})
is smaller than $1/4$).
Each of these closed trajectories is stable
with respect to small perturbations of the initial data:
There is a neighboring closed trajectory passing through
the point that corresponds to the perturbed initial data.

Let us now analyze system (\ref{kg-breather-linearized-nopi}).
We can rewrite it as
$
\dot\Phi={\bf J} {\bf H}\Phi,
$
where
\[\Phi=\left[\begin{array}{c}\phi\\\psi\end{array}\right],
\quad
{\bf J}=\left[\begin{array}{cc}0&1\\-1&0\end{array}\right],
\quad
{\bf H}=\left[\begin{array}{cc}H\sp{+}-\frac{\ln g^2}{2}&0\\0&1\end{array}\right],
\]
with $\phi$, $\psi$ orthogonal to $\pi$.
The matrix $H\sp{+}$
(see (\ref{h-plus}) and thereafter)
has eigenvalues
\[
\sigma(H\sp{+})
=\{-1,\Lambda-1,\dots \},
\]
where the only negative eigenvalue
$-1$ corresponds to $\pi$
and the next eigenvalue,
$\Lambda-1$,
is positive.
The term $-\frac{\ln g^2}{2}$
in (\ref{ekgpp})
further shifts the spectrum upwards as long as $\abs{g}<1$.
Therefore, ${\bf H}$ is positive-definite
on the space $\pi\sp\perp\times l\sp 2(\Z)$.
Thus, $\sqrt{{\bf H}}$ is well-defined;
the matrix ${\bf J}{\bf H}$ is similar to
$\sqrt{{\bf H}}\,{\bf J}\sqrt{{\bf H}}$
and hence has purely imaginary spectrum.

This shows that the breathing peakon solutions
$g(t)\pi$
to (\ref{ddu-is-delta-e-comp}) are spectrally stable.

\begin{remark}
This linearized approach to the stability
does not prove the dynamic orbital stability
of the breathing peakons.
The nonlinear terms may transfer
the energy between ``breathing'' oscillations in $\pi$-direction
and the perturbations in the space $\pi\sp\perp$,
pumping the energy from
system (\ref{kg-breather-linearized-pi})
into (\ref{kg-breather-linearized-nopi}).
It is therefore possible that the energy
of the breathing peakon,
after a small perturbation,
would wind up transferred,
partially or completely,
into directions orthogonal to $\pi$
(and then maybe back).
We do not have a satisfactory description of this process, even though
we believe that techniques such as the Hamiltonian dispersive normal
forms of \cite{SW} could be relevant in addressing it. While outside
the scope of the present study, this may be an interesting question
for future investigations.
\end{remark}

\subsection{Numerical Results}

An orbit of frequency $\wp$ for $g(t)$ can be determined by
solving Eq.~(\ref{eqn-on-g}). This can be done through a variety
of methods such as, e.g., a shooting method in real space or using
quadratures. Here we have chosen a Chebyshev quadrature method
in order to integrate the equation.

\begin{figure}
\begin{center}
\begin{tabular}{cc}
\hskip -24pt
    \includegraphics[width=\middlefig]{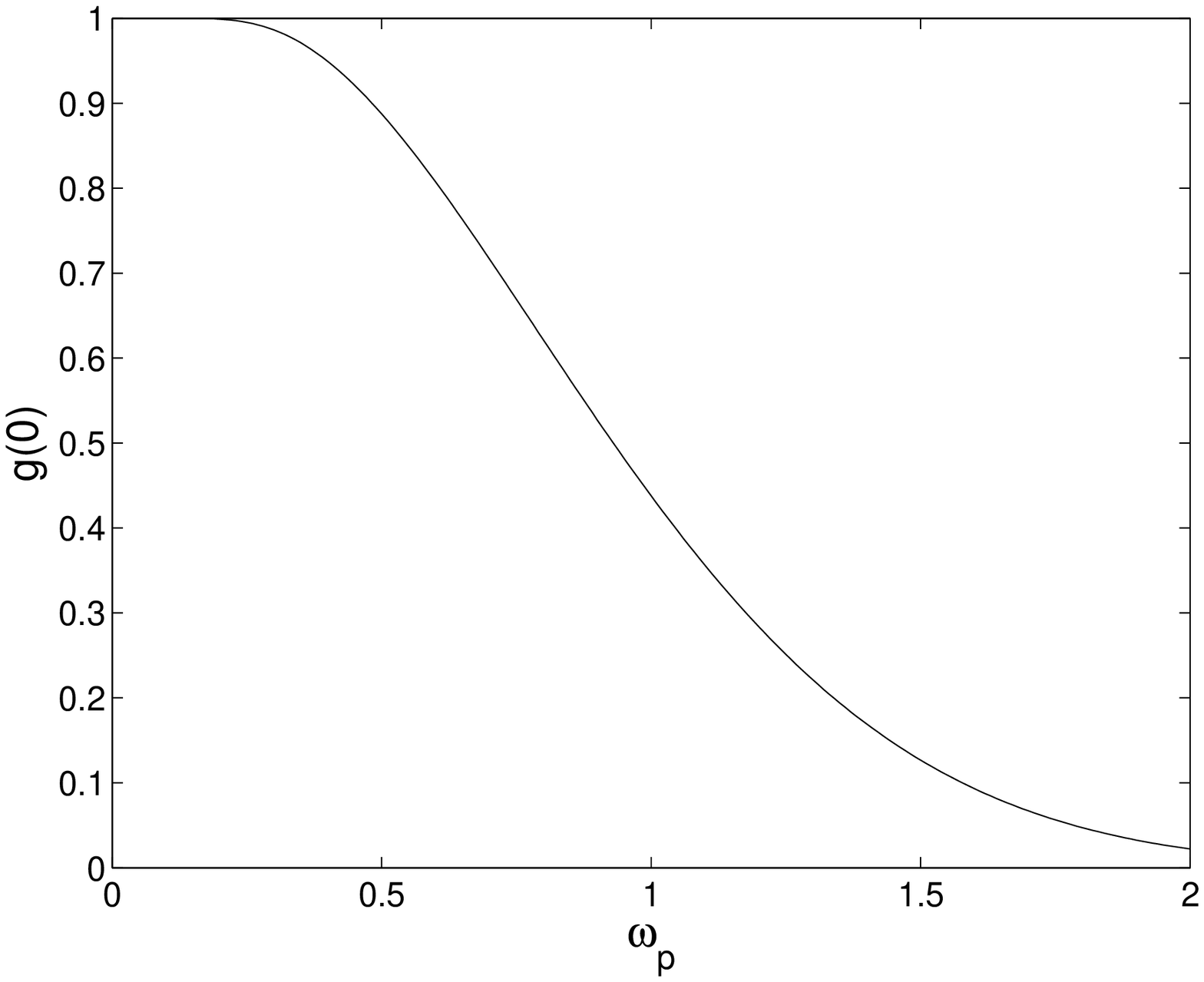} &
    \includegraphics[width=\middlefig]{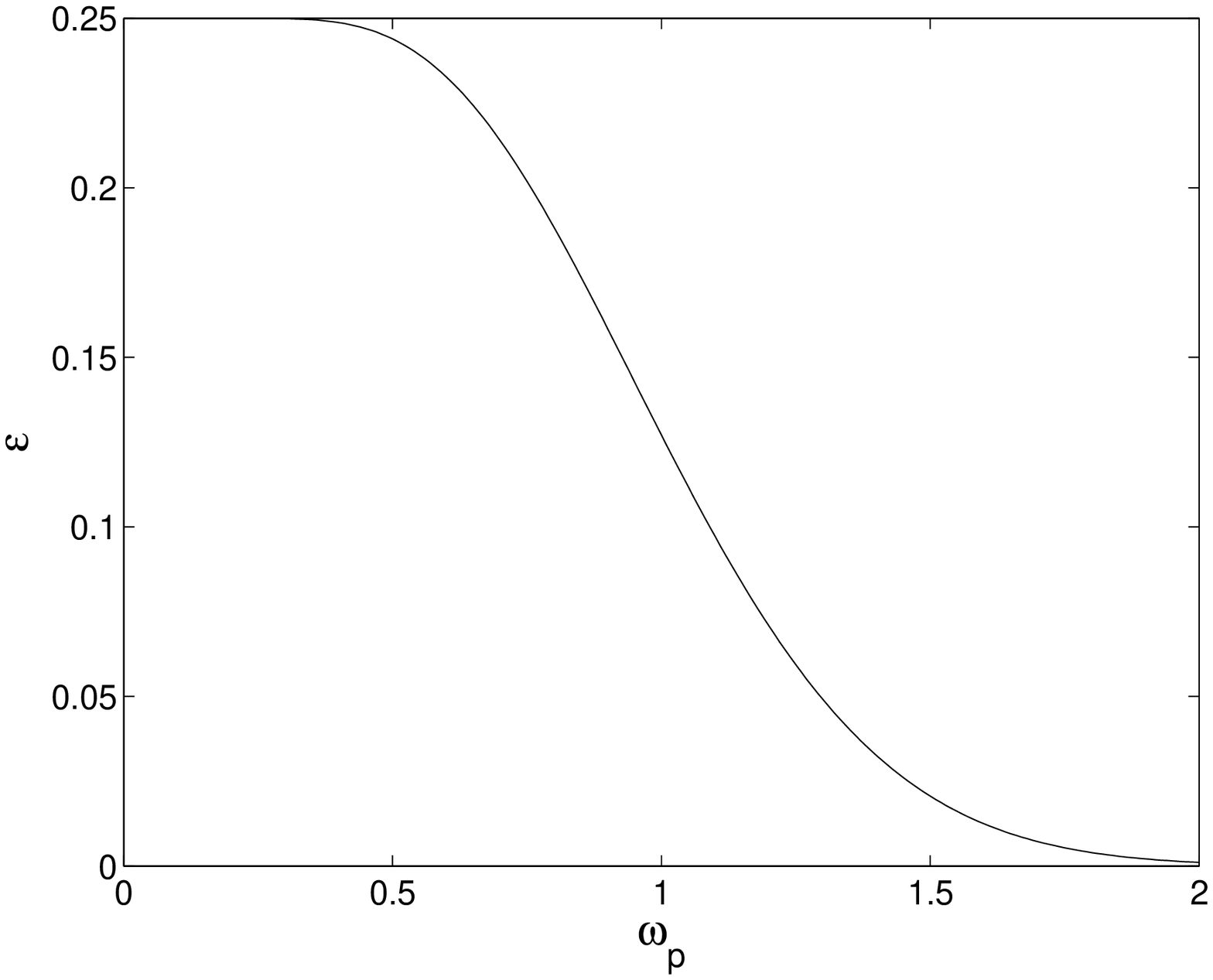}\\
\end{tabular}
\caption{Dependence of $g(0)$ (left) and ${\mathcal E}$ with respect
to the peakon frequency.}
\label{fig:g}
\end{center}
\end{figure}

Figure~\ref{fig:g} shows the dependence of $g(0)$ and ${\mathcal E}$
of the peakon frequency $\wp$. It can clearly be
observed (see the left end of the graphs) that as $\wp \rightarrow
0$, the energy approaches $1/4$ and the amplitude is $1$, hence
the solution is very close to the unstable critical point of
${\mathcal E}=1/4$ and $g=1$.
We can also observe that the peakon frequency can have any value
as there do not exist resonances with the continuous spectrum
(actually, such small
amplitude, extended wave excitations do not exist since
${\mathcal V}''(0)=\infty$).

\begin{figure}
\begin{center}
    \includegraphics[width=\middlefig]{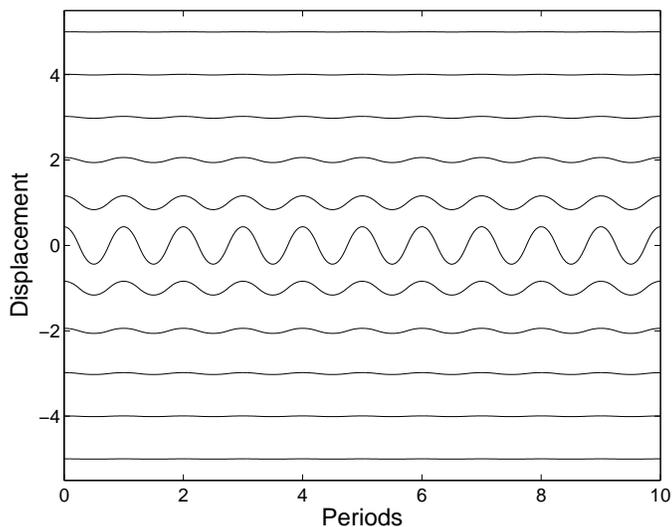}
\caption{Time evolution of the breathing peakon (the displacement
of each of the first few sites is shown as a function of
time).}
\label{fig:evol}
\end{center}
\end{figure}

Figure~\ref{fig:evol} shows the time evolution of a breathing
peakon. It is worth pointing out that the main difference between
DNLS peakons and KG breathing peakons is that, in the first case,
only the first Fourier coefficient is different from zero, whereas
in the second case, there are more non-zero coefficients.

\begin{figure}
\begin{center}
\begin{tabular}{cc}
    \includegraphics[width=\middlefig]{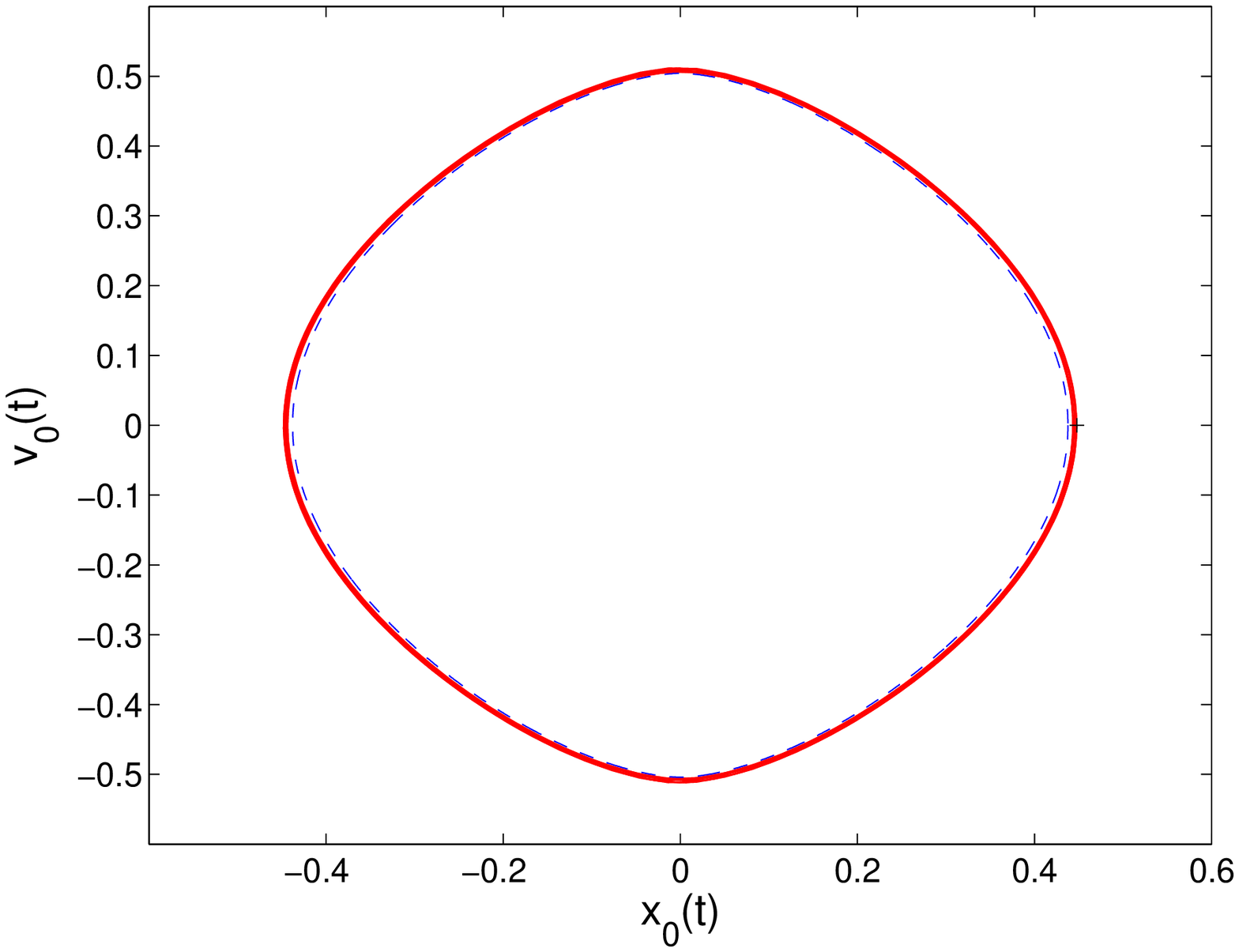} &
    \includegraphics[width=\middlefig]{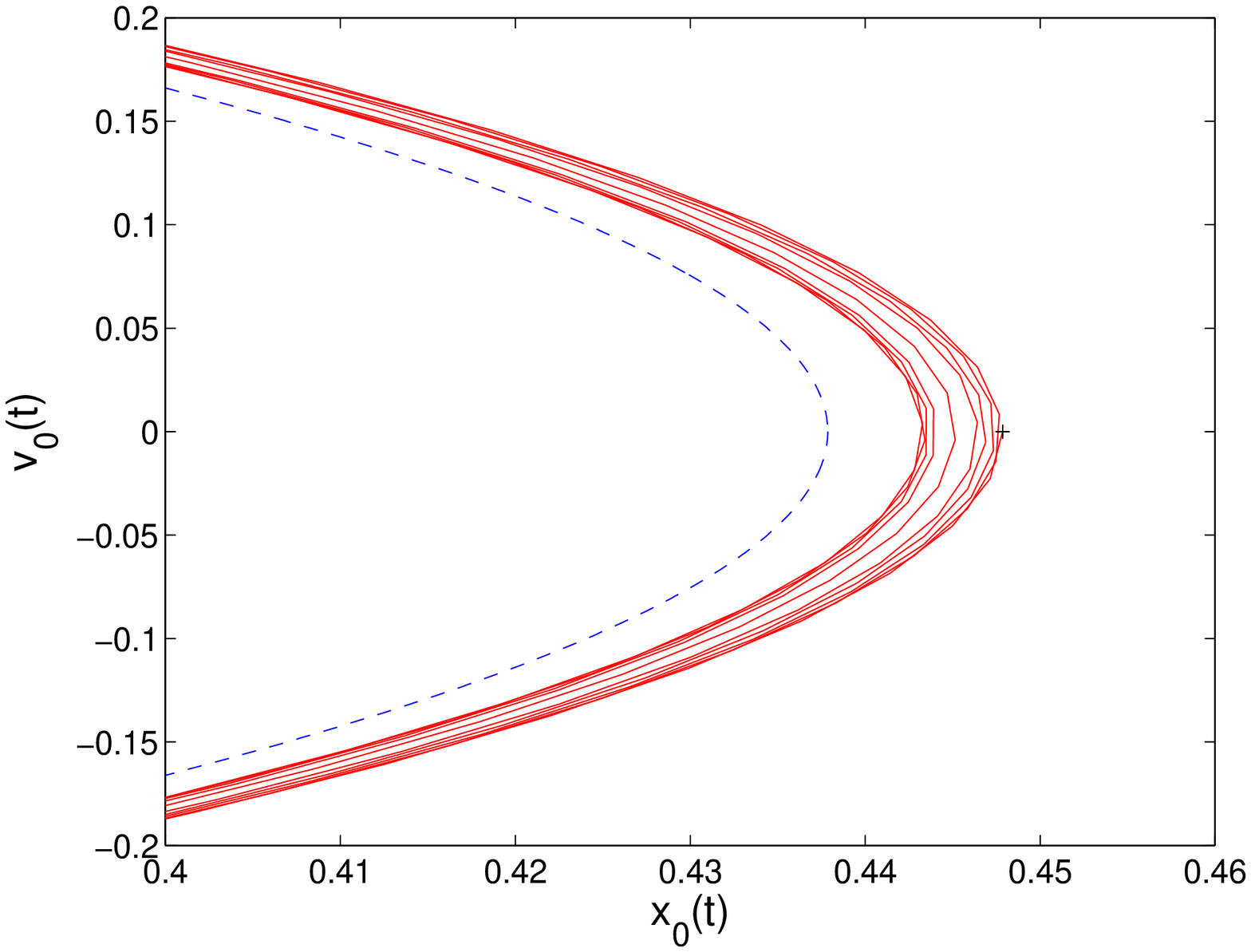} \\
\end{tabular}
\caption{Left panel: Phase space diagram of the central site of a
perturbed (full line) and an unperturbed (dashed line) breathing
peakon with $\wp=1$.
\\
Right panel: Blow-up of the left panel.\\
The cross indicates the initial point of the simulation.}
\label{fig:pert}
\end{center}
\end{figure}

We have confirmed in the numerical simulations that initial
conditions corresponding to a perturbed discrete breathing peakon
stays orbitally close to the exact solution. Figure~\ref{fig:pert}
shows the difference between the evolution of the central particle
of the peakon in a perturbed and an unperturbed case. The
perturbation used is $\xi_n=\varepsilon\delta_{n,0}$ with
$\varepsilon=0.01$. The perturbed peakon appears to be orbitally
stable.

\begin{figure}
\begin{center}
\begin{tabular}{cc}
\hskip -24pt
    \includegraphics[width=\middlefig]{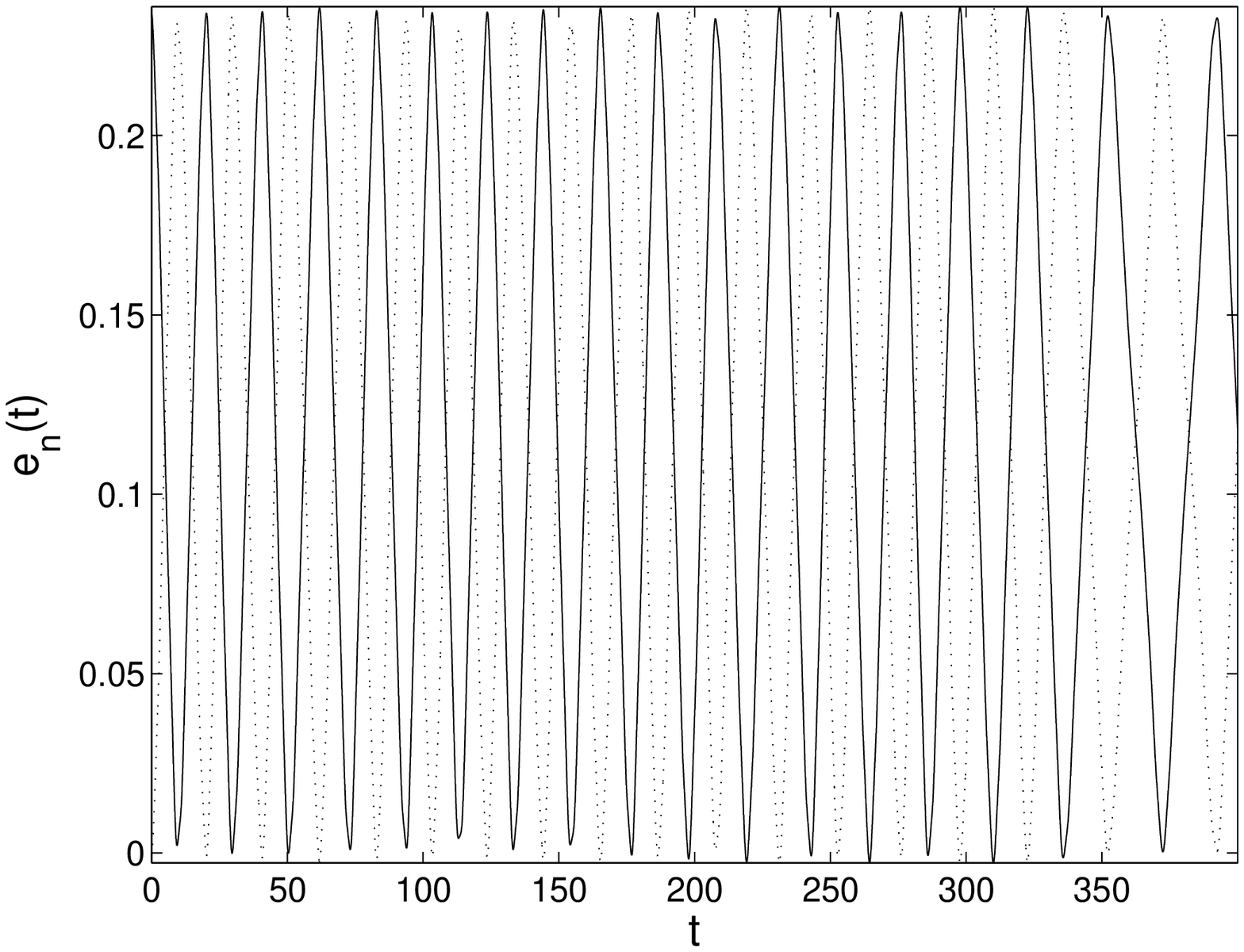} &
    \includegraphics[width=\middlefig]{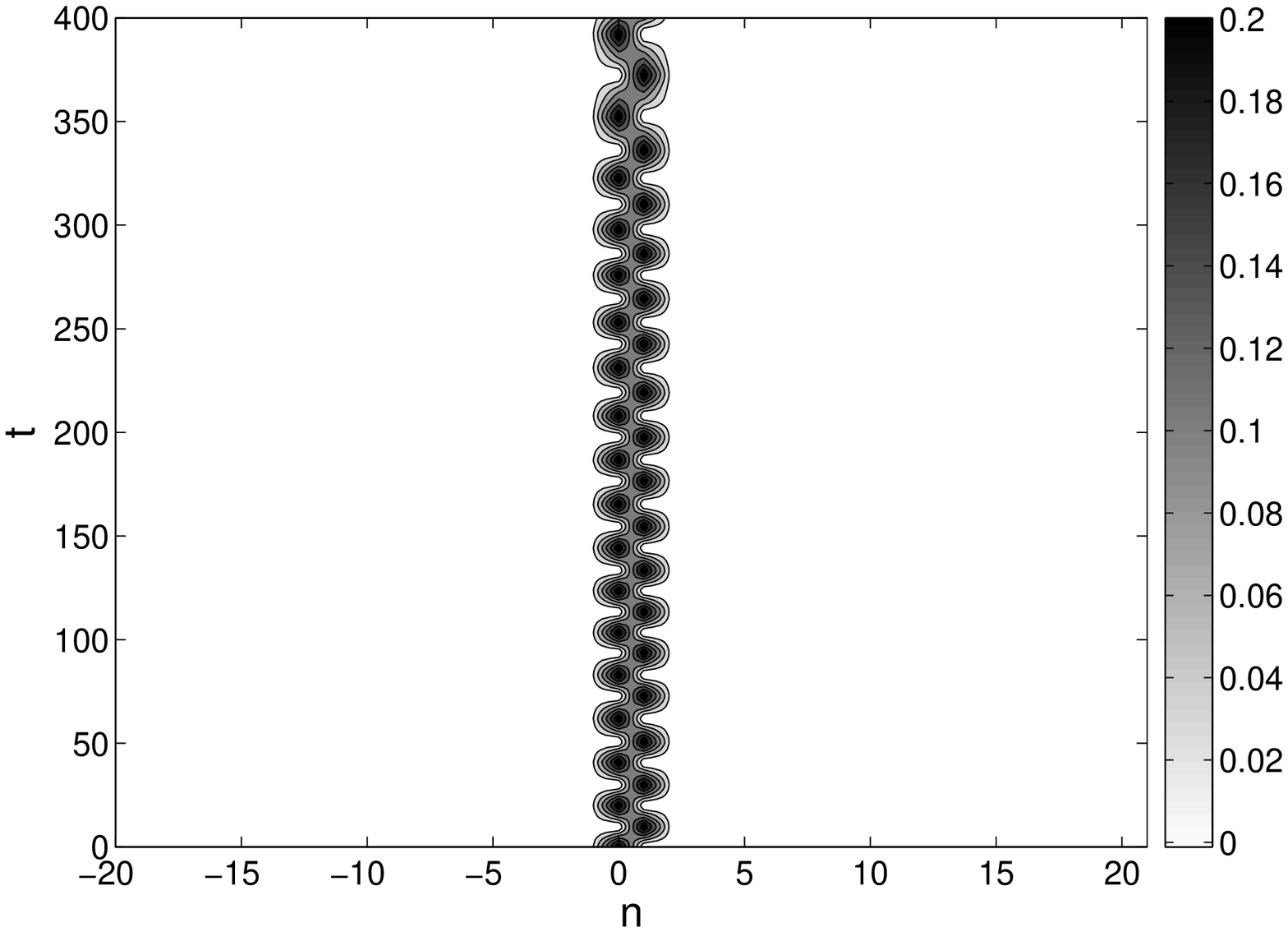}\\
\end{tabular}
\caption{Time evolution of the two-site breathing peakon chain
(with $a=1$ and $\omega_p=1$) perturbed with the antisymmetric
mode.
\\
Left panel: The full line represents $e_0$, the dashed line
represents $e_1$,
where $e_n$ is the energy density of the $n$-th site.
\\
Right panel: Energy density in the space-time evolution.}%
\label{fig:kg2s}
\end{center}
\end{figure}

Contrary to the static case, breathing two-site peakons are not
destroyed by perturbations. Instead, the energy density oscillates
as shown in  Fig.~\ref{fig:kg2s}.

\section{Discussion}
\label{sec:discussion}

In this paper, we have engineered a mathematical model that has
discrete peakons as exact solutions. The continuum version of the
model was also given and its ability to support the continuum
analog of the solutions was highlighted. Both the dispersive
and the nonlinear part of the relevant interactions were connected
to earlier works.
Furthermore, it was advocated that for suitable choice of
the interaction range, this can be a relevant model
in nonlinear optics
with the characteristic
features of photorefractive materials \cite{christo}, while being a
lattice dynamical model of interest in its own right.

We have identified the discrete peakon solutions
analogous to their (discontinuous in the first derivative) continuum
limit, i.e., $\pi\sb n \sim \exp(-|n|)$, as well as their
inter-site siblings. However, a natural question of
interest would be whether there is a more general way of defining
such  solutions in the discrete setting.
This is particularly relevant as solutions
similar to the ones obtained here have appeared in other
contexts. Such examples consist of,
e.g., the waveform of Fig. 1 in \cite{kovalev} (arising from the
presence of an impurity) or that of Fig. 7 in \cite{gaid6}
(arising because of the interplay of nonlinearity with long
range interactions, as is the case in this paper). Perhaps, an alternative
criterion possibly involving the sign of the second difference
close to the center of the wave could be used for a more general
definition of the discrete peakon. This would be an interesting
topic for future studies.

We have also investigated the stability of such discrete
waves and have found discrete peakons to be particularly
interesting from this aspect as well.
We have numerically observed
that in the Klein-Gordon lattice setting such solutions
are always unstable; however the negative energy
direction responsible for this instability is eliminated due to an
additional symmetry (the phase invariance that leads to the
$l\sp 2$-norm conservation) in the case of the DNLS chain.
We showed (using a Derrick-type argument)
that the peakon is not a local minimizer of the energy
in the Klein-Gordon case
and moreover indicated the direction
of the perturbation that leads to a linear instability.
We also showed that in the $U(1)$-invariant DNLS equation
the peakon is a local minimizer
of the energy under the charge constraint and hence is orbitally stable.
Contrary to their one-site counterparts,
two-site peakons proved to be unstable,
as was explicitly demonstrated via
a well-known functional analytic criterion relevant to DNLS type
equations.

Finally, using a
separation of variables approach, we were able to show
the existence of the {\it exact} periodic (breathing)
peakon solutions in the Klein-Gordon lattice.
The breathing peakons were shown to
correspond to the subcritical initial conditions,
while supercritical
initial conditions lead to the amplitude of the
solution tending to infinity (in infinite time),
with the unstable static peakon
being the separatrix between the two types of behavior.
We also systematically tackled the initial
value problem for the Klein-Gordon lattice
and showed that the finite-time blow-up
of the $l\sp 2$-norm of the solution is not possible,
so that the system is globally well-posed.
We then proved the absence of the linear instability
of the breathing peakons.
Yet, the question of the long-time behavior
of perturbed breathing peakons remains open.

It may be interesting to try to extend this class of models to
higher dimensional settings and observe how their dynamical
behavior is affected by the dimensionality of the underlying
lattice.

\vspace{5mm}

AC was supported in part by the National Science Foundation grant
DMS-0200880 and by the Max Planck Institute, Leipzig. JC
acknowledges an FPDI grant from `La Junta de Andaluc\'{\i}a' and
partial support under the European Commission RTN project LOCNET,
HPRN-CT-1999-00163 and the MECD/FEDER project FIS2004-01183. PGK
gratefully acknowledges support from NSF-DMS-0204585, the Eppley
Foundation for Research and from an NSF-CAREER award. We are
thankful to Sergej Flach for bringing Ref. \cite{flach} to our
attention. We are also indebted to an anonymous referee for bringing
to our attention a recent paper \cite{oster}, as well as several
earlier relevant works.


\end{document}